\documentclass[usenatbib]{mn2e}
\usepackage{multirow}
\usepackage{rotating}
\usepackage{colortbl}
\usepackage{color}
\usepackage[fleqn]{amsmath}
\usepackage{amssymb}
\usepackage{amsfonts}
\usepackage{verbatim}
\usepackage{scalefnt}
\usepackage[percent]{overpic}

\newcommand{\apjl}{ApJ}
\newcommand{\apjs}{ApJS}
\newcommand{\ao}{Appl.~Opt.}
\newcommand{\aap}{A\&A}

\newcommand{\beq}{\begin{equation}}
\newcommand{\eeq}{\end{equation}}

\definecolor{grey}{rgb}{0.5,0.6,0.7}
\def \simlt { \lower .75ex \hbox{$\sim$} \llap{\raise .27ex \hbox{$<$}} }
\definecolor{purple}{rgb}{0.65,0.15,0.9}
\definecolor{darkorange}{rgb}{0.8,0.3,0}
\definecolor{olive}{rgb}{0.4,0.6,0.25}
\definecolor{darkgreen}{rgb}{0,0.7,0}
\definecolor{darkred}{rgb}{0.5,0,0}

\title[Galaxy number counts around high-redshift massive black holes]{The diverse galaxy counts in the environment of high-redshift massive black holes in Horizon-AGN}               
\author[Habouzit et al.]{M\'{e}lanie Habouzit$^{1}$\thanks{E-mail: mhabouzit@flatironinstitute.org},
Marta Volonteri$^{2}$,
Rachel S. Somerville$^{1,3}$,
Yohan Dubois$^{2}$,
\newauthor
S\'{e}bastien Peirani$^{2,4}$,
Christophe Pichon$^{2,6,7}$,
Julien Devriendt$^{5}$\\
$^1$ Center for Computational Astrophysics, Flatiron Institute, 162 5th avenue, 10010 NY, USA\\
$^2$ Institut d'Astrophysique de Paris, Sorbonne Universit\'{e}s, CNRS, UMR 7095, 98 bis bd Arago, 75014 Paris, France\\
$^3$ Department of Physics and Astronomy, Rutgers University, 136 Frelinghuysen Rd, Piscataway, NY 08854, USA\\
$^4$ Observatoire de la C\^ote d'Azur, CNRS, Laboratoire Lagrange, Bd de l'Observatoire, CS 34229, F-06304 Nice Cedex 4, France\\
$^5$ Subdepartment of Astrophysics, University of Oxford, Keble Road, Oxford, OX1 3RH, UK\\
$^6$ School of Physics, Korea Institute for Advances Study (KIAS), 85 Hoegiro, Dongdaemun-gu, Seoul, 02455, Republic of Korea\\
$^7$ Institute for Astronomy, University of Edinburgh, Royal Observatory, Blackford Hill, Edinburgh, EH9 3HJ, United Kingdom\\
}
\begin{document}
\maketitle

\begin{abstract}

High-redshift quasars are believed to reside in highly biased regions of the Universe, where black hole growth is sustained by an enhanced number of mergers and by being at the intersection of filaments bringing fresh gas. This assumption should be supported by an enhancement of the number counts of galaxies in the field of view of quasars. While the current observations of quasar environments do not lead to a consensus on a possible excess of galaxies, the future missions JWST, WFIRST, and Euclid will provide new insights on quasar environments, and will substantially increase the number of study-cases. We are in a crucial period, where we need to both understand the current observations and predict how upcoming missions will improve our understanding of BH environments. Using the large-scale simulation Horizon-AGN, we find that statistically the most massive BHs reside in environments with the largest galaxy number counts. However, we find a large variance in galaxy number counts, and some massive BHs do not show enhanced counts in their neighborhood. Interestingly, some massive BHs have a very close galaxy companion but no further enhancement at larger scales, in agreement with recent observations.
We find that AGN feedback in the surrounding galaxies is able to decrease their luminosity and stellar mass, and therefore to make them un-observable when using restrictive galaxy selection criteria. Radiation from the quasars can spread over large distances, which could affect the formation history of surrounding galaxies, but a careful analysis of these processes requires radiative transfer simulations.

\end{abstract}


\section{Introduction}
\label{sec:intro}

Quasars, powered by supermassive black holes (BHs), are among the brightest objects in the Universe.
So far, we have observed about $\sim 200$ quasars at $z>5$ \citep[][and references therein]{Fan2006,Willot2010,Mortlock2011,Venemans2013}. 
These very high-redshift BHs are on the tail of the mass distribution, and are already as massive as the most massive BHs in nearby galaxies. 
The presence of a few of these extreme BHs when the Universe was less than 1 Gyr \citep[$z\sim 7$,][]{Mortlock2011,2018ApJ...856L..25B} offers us the tightest constraint that we have on the formation of supermassive BHs.
BHs must have formed at very early times and grown significantly over the first billion years of the Universe. The population of high-redshift quasars is therefore a formidable laboratory to investigate their early growth and constrain theoretical models of accretion mechanisms.
The properties of the quasar host galaxies are also fascinating, as they are already chemically evolved, dusty and metal-rich \citep[see the recent review by][on the formation and galaxy host properties of quasars]{2017PASA...34...31V}. 

One of the mostly debated issues related to the discovery of distant quasars is whether they are embedded in overdense regions of the Universe. 
Extrapolating relations between BH mass and their host galaxy properties \citep[e.g.,][]{2013ARA&A..51..511K}, one would argue that these bright quasars reside in the most massive dark matter haloes that have collapsed at these epochs (of about $M_{\rm h}\sim 10^{13} M_{\odot}$) \citep[e.g.][]{Ferrarese2002,Wyithe2003,Fanetal2004}, the same haloes around which more galaxies than average are believed to reside \citep{KauffmannHaehnelt02,2009MNRAS.394..577O}. Theoretically, galaxy clustering around high-redshift quasars can be defined by the quasar-galaxy cross-correlation, i.e. as the excess of the galaxy number counts in the quasar field of view compared to an average density of galaxies. In the case of non-clustering, the number of galaxies in the observed field should simply be proportional to the observed volume.

During the last decade, an increasing number of observational campaigns aiming at investigating the galaxy clustering around such massive BHs, via number counts of galaxies, have been conducted \citep[e.g.][]{2005ApJ...622L...1S,Willotetal2005,2006ApJ...640..574Z,2009ApJ...695..809K,2010ApJ...721.1680U,2013MNRAS.432.2869H,2014MNRAS.442.3454S,2014A&A...568A...1M,2014AJ....148...73M,2017ApJ...834...83M}. 
Despite the fact that many observational efforts have been devoted to answer the question of the environments of high-redshift quasars, we have not reached a consensus yet. 

\begin{figure*}
\centering
\includegraphics[scale=0.65]{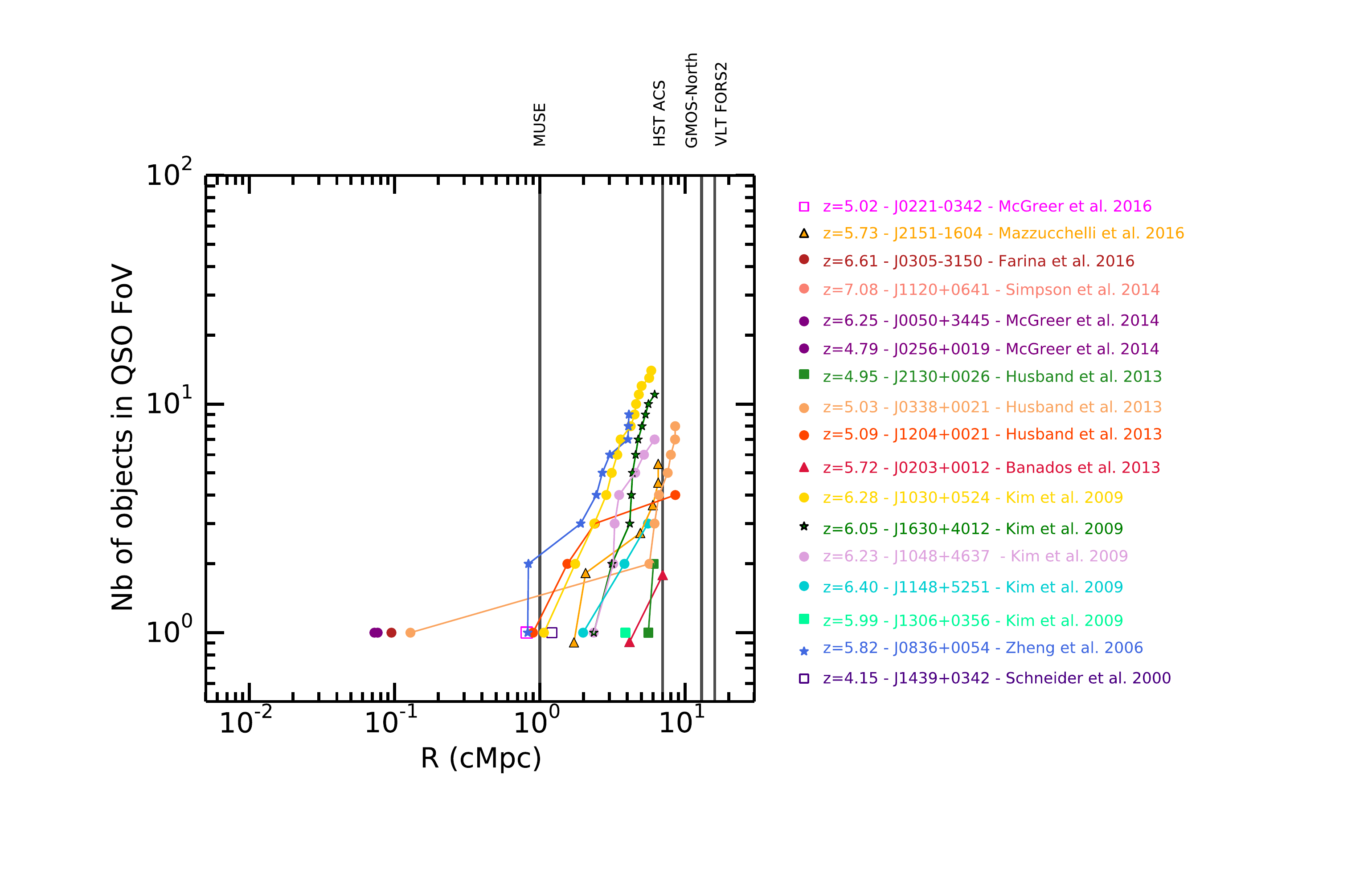}
\caption{Number of objects (i.e. galaxies, or other quasars) detected in the field of view of high-redshift quasars. Single symbols represent the observation of one single object in the quasar field of view. Solid lines represent the cumulative number of objects observed around a given quasar. We show the particular cases of the quasar J0221-0342 \citep{2016AJ....151...61M}, J2151-1604 \citep{2017ApJ...834...83M}, J0305-3150 (Farina et al. 2017), J1120+0641 \citep{2014MNRAS.442.3454S}, J0050+3445 \citep{2014AJ....148...73M}, J0256+0019 \citep{2014AJ....148...73M}, J2130+0026 \citep{2013MNRAS.432.2869H}, J0338+0021 \citep{2013MNRAS.432.2869H}, J1204+0021 \citep{2013MNRAS.432.2869H}, J0203+0012 \citep{2013ApJ...773..178B}, J1030+0524 \citep{2009ApJ...695..809K,2005ApJ...622L...1S,Willotetal2005,2014A&A...568A...1M}, J1630+4012 \citep{2009ApJ...695..809K}, J1048+4637 \citep{2009ApJ...695..809K,Willotetal2005,2014A&A...568A...1M}, J1148+5251 \citep{2009ApJ...695..809K,Willotetal2005,2014A&A...568A...1M}, J1306+0356 \citep{2009ApJ...695..809K}, J0836+0054 \citep{2006ApJ...640..574Z}, J1439+0342 \citep{2000AJ....120.2183S}. 
We have computed the separation distances with the cosmological parameters that have been used for the simulation Horizon-AGN. 
The distances or FoV probed by several instruments (MUSE, HST ACS, GMOS-North, and VLT FORS2) at $z\sim6$ are indicated as black vertical lines, the FoV of LBT LBC and Subaru Suprime Cam, are larger and of about $\sim 50-70 \, \rm cMpc$.}
\label{fig:obs_data}
\end{figure*}

To illustrate the current puzzle set by the $\sim 20$ observations of quasar environments available so far, we show in Fig.~\ref{fig:obs_data} the number counts of objects in the field of these quasars. The figure shows the number counts of objects as a function of their projected distance from the quasars. Different colors indicate different high-redshift quasars, which are listed on the right side of the figure. The grey vertical lines give the sizes of the fields of view that are probed by the current instruments MUSE, HST ACS, GMOS-North, and VLT FORS2.  
The environment of some quasars are very clustered, i.e. overdense, presenting an excess of the number counts of galaxies in the quasar field of view compared to a control field, and some quasars have a very close galaxy companion (or a quasar companions for two of them), but some other fields do not show an excess of galaxies.
The observational approaches used to measure a possible enhancement of galaxy counts in the quasar's environment rely on the detection of Lyman Break Galaxies (LBGs), or Lyman Alpha Emitters (LAEs). The first method is based on the detection of a drop in the galaxy spectrum when going through different color bands. 
The second method relies on the detection of bright Ly$\alpha$ emission.
While the LBG detections suffer from large redshift uncertainty with $\Delta z \sim 1$, detections of LAEs allow for a better estimation of redshift with $\Delta z\sim 0.1$, but could also suffer from uncertain biases \citep{2011ApJ...726...38Z,2017arXiv171006171B}.
Observing galaxies at high redshift, $z\geqslant 5$, in quasar fields of view is very challenging. Current studies and facilities suffer from redshift uncertainties, from probing only small fields of view around the quasars, reduced capacity to observe faint galaxies, and unknown selection biases.  
We provide a more detailed review of the $\sim 20$ observations in Appendix~\ref{sec:observations}, with a discussion on the uncertainties of the galaxy number count measurements in the quasar fields of view, with a particular emphasis on the possible underlying reasons explaining the differences of the observational findings.

Upcoming observational missions will have the potential to shed new light on the environments of high-redshift quasars. They will especially increase our ability to observe fainter galaxies, and larger fields of view. The upcoming James Webb Space Telescope (JWST) will cover the wavelength range 0.6 to 5.0 $\micron$ with the NIRCam instrument, and a field of view of $\sim 10 \, \rm arcmin^2$, which corresponds to a size of field in the range $\sim 1-1.3\, \rm Mpc$ for $z\sim 4-6$. The high resolution of JWST, which should be able to resolve galaxies with stellar mass of about $M_{\star} \geqslant 10^{7}\, \rm M_{\odot}$ in these redshift range \citep{2018ApJS..236...33W,2017jwst.prop.1345F}, will be crucial to observe or re-observe small regions around high-redshift quasars where faint galaxies could have been missed in previous campaigns. 
Other missions such as WFIRST and Euclid will probe larger fields of view. WFIRST should be able to resolve galaxies of $M_{\star}\geqslant 10^{8}\, \rm M_{\odot}$ \citep{2015arXiv150303757S} with a field of view of $\sim 1000 \, \rm arcmin^{2}$, which corresponds to $\sim 11-13\, \rm Mpc$. Euclid will probe even larger fields of view at the expense of resolution, and should be able to detect galaxies of $M_{\star}\geqslant 10^{9.5}\, \rm M_{\odot}$ \citep{2016SPIE.9904E..0OR}.

We are in a crucial period. While the current observations of quasar environments do not lead to a consensus on a possible excess of galaxies, future missions will provide new insights about the environment of high-redshift quasars, and will substantially increase the number of study-cases.
It is critical to assess both whether the most massive high-redshift BHs statistically reside in overdense regions of the Universe, but perhaps more importantly to assess what is the diversity of high-redshift BH environments we can expect theoretically. 
The goal of the present paper is to provide theoretical predictions on the galaxy number counts around massive BHs in order to  understand both the intriguing diversity of quasar environments found in observations, and how the next generation missions such as JWST, WFIRST or Euclid, could improve our understanding of massive BH environments.\\

Hydrodynamical cosmological simulations offer us a key tool to study self-consistently the evolution of both galaxies and their massive BHs. Unfortunately, theoretically the environment of high-redshift quasars could only be assessed in very large simulations because of their low observed number density of $\sim 1 \rm{Gpc^{-3}}$. For comparison, the largest hydrodynamical cosmological simulation to date, BlueTides, has a side length of $400 h^{-1}\rm cMpc$ \citep{2017MNRAS.467.4243D}.  
Previous theoretical studies have tried to overcome the limited-volume issue of cosmological simulations, by either using dark matter only simulations with a large volume, or using zoom-in simulations of massive haloes. \citet{2009MNRAS.394..577O} use the dark matter simulation Millenium \citep{2005Natur.435..629S} (with a box side length of $500 h^{-1}\rm Mpc$), and a semi-analytical galaxy formation model to paint galaxies on top of the simulation \citep{2007MNRAS.375....2D}, from which mock surveys have been derived. They show that the common assumption of quasars being embedded in the most massive dark matter haloes can not be ruled out, nor confirmed, and that the galaxy density enhancement can actually extend to large scale of several tens of $\rm Mpc$. 
In the BlueTides simulation, the most massive BHs are not found in haloes with the largest overdensities but instead in haloes with the lowest tidal fields, i.e. more radial flows of cosmic gas \citep{2017MNRAS.467.4243D}.
\citet{2018MNRAS.474..597T} have recently used the same simulation to study the descendants of the simulated high-redshift quasars. 
\citet{Costa:2013aia} studied the environment of bright quasars using re-simulations extracted from the large scale simulation Millennium, at $z\sim 6$. Based on synthesized rest-frame UV images for the simulation (comparable with those of the Hubble Space Telescope HST), they compare the galaxy luminosity function of the simulation to the one derived observationally, for different models of galactic winds implementation. They find that strong supernova (SN) winds are able to delay the early BH growth, and provide more fuel to feed BH growth at later times. Other simulations have shown that BH growth is stunted by SN feedback at early times \citep{2015MNRAS.452.1502D,2017MNRAS.468.3935H}. In these simulations, SN feedback strongly impacts the dense gas, while in \citet{Costa:2013aia} the dynamics of winds is decoupled from that of the dense star-forming gas.
In their simulations, SN winds can favor the growth of BHs indirectly, i.e. galactic winds from satellite galaxies can eject gas and push matter toward the potential well of the main halo, which results in the feeding of the central BHs of this latter halo. 
The impact of AGN feedback is also investigated and is found to not have a strong effect on the number of galaxies in the field, though this investigation only focused on few zoom-in simulations. 
These theoretical predictions have been focussing so far on either large dark matter only simulations with semi-analytic models for which the physics of BHs is not followed self-consistently, or on analyzing in details few zoom-in simulations and realizing mock observations matching a particular instrument capacities such as HST.  \\

In this paper we use the state-of-the-art simulations Horizon-AGN \citep{2014MNRAS.444.1453D} and Horizon-noAGN \citep{2017MNRAS.472.2153P} to analyze the number counts of galaxies around the most massive BHs. 
The simulations both use the same initial conditions, and share the exact same physics, except that there are no BHs and therefore no AGN feedback in Horizon-noAGN. 
The twin simulations have been able to reproduce a large range of galaxy properties \citep[e.g.][]{2014MNRAS.444.1453D,2017MNRAS.472.2153P,2017MNRAS.tmp..224K}, as well as BH properties \citep{2016MNRAS.460.2979V}. The simulations are therefore well suited for the present investigation. As the volume of Horizon-AGN is $(100 \, \rm cMpc)^{3}$, we are not likely to find any BHs as massive as the most extreme quasars with $M_{\rm BH}\geqslant 10^{9}\, M_{\odot}$.
Typical bolometric luminosities for these extreme objects are $L_{\rm bol}\geqslant 10^{47}\, \rm erg/s$\footnote{Quasar masses are estimated from the virial theorem, i.e. that their masses are determined as a function of velocity dispersion and the sizes of their broad-line regions, which is measured from line widths.}. However, the simulation is still able to produce the low-mass end of the observed quasars, with a few BHs with $M_{\rm BH}\geqslant 10^{8}\, \rm M_{\odot}$ at $z\sim 5$. In observations if we assume an Eddington luminosity for these BHs, they would have bolometric luminosities of a few $10^{46}\, \rm erg/s$. Today, we start observing a regime of {\it fainter} high redshift quasars with estimated BH mass of $M_{\rm BH}\geqslant 10^{7.6}\, \rm M_{\odot}$ \citep[][for quasars at $6.1<z<6.7$]{2019arXiv190407278O}. With our set of simulations we probe this regime. In this paper we study the galaxy number counts in the field of view of the massive BHs (we do not use the AGN luminosity as a criterion). Current observations of quasars are likely to be biased towards luminous quasars (particularly at high redshift), and likely miss a population of BHs, as massive as the observed ones, but with lower accretion rate and so luminosity. Here, we consider the full population of the massive BHs to study their environments.
We describe the two simulations and their sub-grid physics models in Section 2.
We aim at understanding whether the most massive BHs are embedded in denser regions of the Universe, and whether there is a scatter in the number counts of galaxies around supermassive BHs at high redshift. 
Direct comparisons with the $\sim 20$ available observations of quasar environments is difficult as the galaxy selections are very heterogeneous (e.g., different redshift uncertainty, observed surface on the sky, galaxy selection criteria). A review of the state of the field is provided in Appendix A, and the different selections and instruments used for these observations are listed in Table A1.  
In Section 3, we study the differences between 3D spherical number counts and projected 2D counts. In Section 4, we analyze the galaxy 2D projected number counts around massive BHs, quantify the diversity of high-redshift BH environments, study how redshift uncertainties can affect the signal, and whether the galaxy number count is a good tracer of dark matter halo mass.
In Section 5, we use the simulation Horizon-noAGN \citep{2017MNRAS.472.2153P} to test whether AGN feedback can suppress the enhancement of galaxy counts in the field of view of these BHs.

\section{Simulation parameters}
\subsection{{\sc Ramses} code and initial conditions}
\label{subsec:ramses}
In this paper we analyse two large scale state-of-the-art cosmological hydrodynamical simulations, Horizon-AGN \citep{2014MNRAS.444.1453D} and Horizon-noAGN \citep{2017MNRAS.472.2153P}. A full description of the simulation Horizon-AGN can be found in \citet{2014MNRAS.444.1453D}.
The simulations are run using the adaptive mesh refinement hydrodynamical cosmological code {\sc Ramses } \citep{2002A&A...385..337T}. 
Particles are projected on the grid with a Cloud-In-Cell interpolation and the Poisson equation is solved with an adaptive Particle-Mesh solver.
The Euler equations are solved with a second-order unsplit Godunov scheme using an approximate HLLC Riemann solver, with a Min-Mod total variation diminishing scheme to interpolate the cell-centered values to their edge locations.
The initial mesh is refined with seven levels of refinement, which corresponds to a spatial resolution up to $\Delta x=1\, \rm kpc$.
Cells are refined (unrefined) based on a quasi-Lagrangian criterion: with more (less) than 8 DM particles in a cell, or with a total baryonic mass higher (smaller) than 8 times the DM mass resolution. To keep the refinement homogeneous in physical units throughout cosmic time a new refinement level is allowed only when the expansion factor is doubled, namely for $a_{\rm exp}=0.1,0.2,0.4$ and so on.\\
The two simulations use a standard $\Lambda$ cold dark matter cosmology, with parameters compatible with those of the {\it Wilkinson Microwave Anisotropy Probe} (WMAP7) parameters \footnote{These cosmological parameters are compatible within 10 per cent relative variation with those of {\it Planck} parameters \citep{planck2013}.} \citep{Spergel2007}: total matter density $\Omega_{m}=0.272$, dark matter energy density $\Omega_{\Lambda}=0.728$, amplitude of the matter power spectrum $\sigma_{8}=0.81$, spectral index $\rm{n_{s}}=0.967$, baryon density $\Omega_{b}=0.045$ and Hubble constant $\rm{H_{0}}= 70.4\, \rm{km\,  s^{-1} \, Mpc^{-1}}$.

\subsection{Physics of the simulation}
Here, we summarize the main sub-grid models of the simulations, all details can be found in \citet{2014MNRAS.444.1453D}.\\

\noindent {\bf Radiative cooling and photoheating}\\
Radiative cooling is modeled with the cooling curves of \cite{1993ApJS...88..253S}, the gas cools down to $10^{4}\, \rm K$ through H, He, with a contribution from metals. 
To mimic reionization, photoheating from an uniform ultraviolet radiation background is added, taking place after redshift z=10, following \citet{1996ApJ...461...20H}.
Metallicity of the gas is modeled as a passive variable, which makes it easily trackable over the gas flow through cosmic evolution. 
An initial metallicity background of ${\rm Z}=10^{-3}\, \rm Z_{\odot}$ is used for the two simulations. The metallicity is then modified by stellar physical processes, such as the injection of gas ejecta from SN explosions, and stellar winds. The simulations account for the release of various chemical elements synthesized in stars and released by stellar winds and SNe: O, Fe, C, N, Mg and Si, however these elements do not contribute separately to the cooling curve. \\

\noindent {\bf Star formation and SN feedback}\\
Star formation occurs in dense ($\rho>\rho_{0}$, with $\rho$ the density of the gas, $\rho_{0}=0.1\, \rm{H\, cm}^{-3}$ the gas hydrogen number density threshold, see equation~\ref{eq:SF}) and cold ($T<T_{0}$, see equation~\ref{eq:TSF}) cells. The value $\rho_{0}$ is computed in order to resolve the Jeans mass with 4 cells \citep{1997ApJ...489L.179T}.
Star formation is modeled with a Kennicutt-Schmidt law:
\begin{equation}
\frac{d\rho_{\star}}{dt}=\epsilon_{\star}\frac{\rho}{t_{\rm{ff}}},
\label{eq:SF}
\end{equation}
with $\dot{\rho_{\star}}$ the star formation rate density, $\epsilon_{\star}=0.02$ the star formation efficiency (constant with redshift), $\rho$ the density of the gas and $t_{\rm{ff}}$ the free-fall time of the gas. Stars are created with a Poisson random process calculating the probability to form N stars with a mass resolution of  $m_{\rm{res,\star}}=\rho_{0}\Delta x^{3}\sim 2\times 10^{6}\, M_{\rm \odot}$.\\
The gas follows an adiabatic equation-of-state (EoS) for monoatomic gas with adiabatic index  $\gamma=5/3$, except at high gas densities $\rho>\rho_0$, where a polytropic EoS is used to increase the gas pressure in dense gas in order to limit excessive gas fragmentation by mimicking heating of the interstellar medium from stars~\citep{springel&hernquist03}:
\begin{equation}
T=T_{0}\left(\frac{\rho}{\rho_{0}}\right)^{\kappa-1},
\label{eq:TSF}
\end{equation}
with $T$ the gas temperature, $T_{0}$ the temperature threshold, $\rho_{0}$ the density threshold, and $\kappa=4/3$ the polytropic index of the gas. \\
Feedback from Type II SNe, stellar winds, and Type Ia SNe, is implemented, and assuming a Salpeter (1955) initial mass function with a low-mass cut-off of $0.1 \, M_{\rm \odot}$, and a high-mass cut-off of $100 \, M_{\rm\odot}$, and a kinetic feedback as in \citet{2008A&A...477...79D}. A completed description of the model can be found in \citet{2017MNRAS.467.4739K}. \\

\noindent {\bf Black hole formation, fueling and feedback}\\
BHs are represented as collisionless sink particles, that are able to accrete gas from surrounding cells, and merge together. 
BH formation occurs in dense regions where $\rho>\rho_{0}$, initial BH mass is fixed and equal to $M_{\rm seed}=10^{5}\,M_{\rm \odot}$, or less if the amount of gas needed is not available. BHs are not allowed to form closer that $50 \, \rm kpc$ of an existing BH, to avoid the formation of several BHs within the same galaxy. However, several BHs are found in galaxies, because of galaxy-galaxy mergers \citep[see][]{2016MNRAS.460.2979V}.

The accretion onto the BHs is described by the minimum between a Bondi-Hoyle-Lyttleton accretion rate and the Eddington accretion rate, with:
\begin{eqnarray*}
\dot{M}_{\rm BH}=\min(\frac{4\pi \alpha G^{2} M_{\rm BH}^{2}\bar{\rho}}{(\bar{c}^{2}+\bar{v}^{2})^{3/2}},\frac{4\pi G M_{\rm BH} m_{p}}{\epsilon_{r} \sigma_{T} c})
\end{eqnarray*}
where $\alpha$ is a boost factor \citep{Booth2009} ($\alpha=(\rho/\rho_{0})^{2}$ for $\rho>\rho_{0}$, and $\alpha=1$ otherwise), G is the gravitational constant, $M_{\rm BH}$ the mass of the BH, $\bar{\rho}$ the average density of the medium, $\bar{c}$ the average of the sound speed, $\bar{v}$ the average velocity of the gas relative to the BH, $m_{p}$ the proton mass, $\epsilon_{r}=0.1$ the radiative efficiency, and $\sigma_{T}$ the Thomson cross-section.

BHs can impact their surroundings, and particularly their host galaxies, by means of AGN feedback. The simulation Horizon-AGN includes AGN feedback, whereas Horizon-noAGN does not. AGN feedback is a combination of two modes \citep[see][for a complete description of the implementation]{2012MNRAS.420.2662D}: radio mode for low accretion rates when $\dot{M}_{\rm BH}/\dot{M}_{\rm Edd}<0.01$, and quasar mode for high accretion rates with $\dot{M}_{\rm BH}/\dot{M}_{\rm Edd}>0.01$. 
The radio mode deposits AGN feedback energy into a bipolar cylindrical outflow with a jet velocity of $10^{4}\, \rm km s^{-1}$, with an energy of $\dot{E}_{\rm AGN}=\epsilon_{f} \epsilon_{r} \dot{M}_{\rm BH}c^{2}$, with an efficiency $\epsilon_{f}=1$. 
The quasar mode is modeled as an isotropic injection of thermal energy into the gas within a sphere of radius $\Delta x$ with an energy of $\dot{E}_{\rm AGN}=\epsilon_{f} \epsilon_{r} \dot{M}_{\rm BH}c^{2}$, and an efficiency $\epsilon_{f}=0.15$ so that the scaling relations between BH mass and galaxy properties such as velocity dispersion, and BH density in the local Universe \citep{2012MNRAS.420.2662D} are reproduced.

\begin{figure*}
\centering
\includegraphics[scale=0.65]{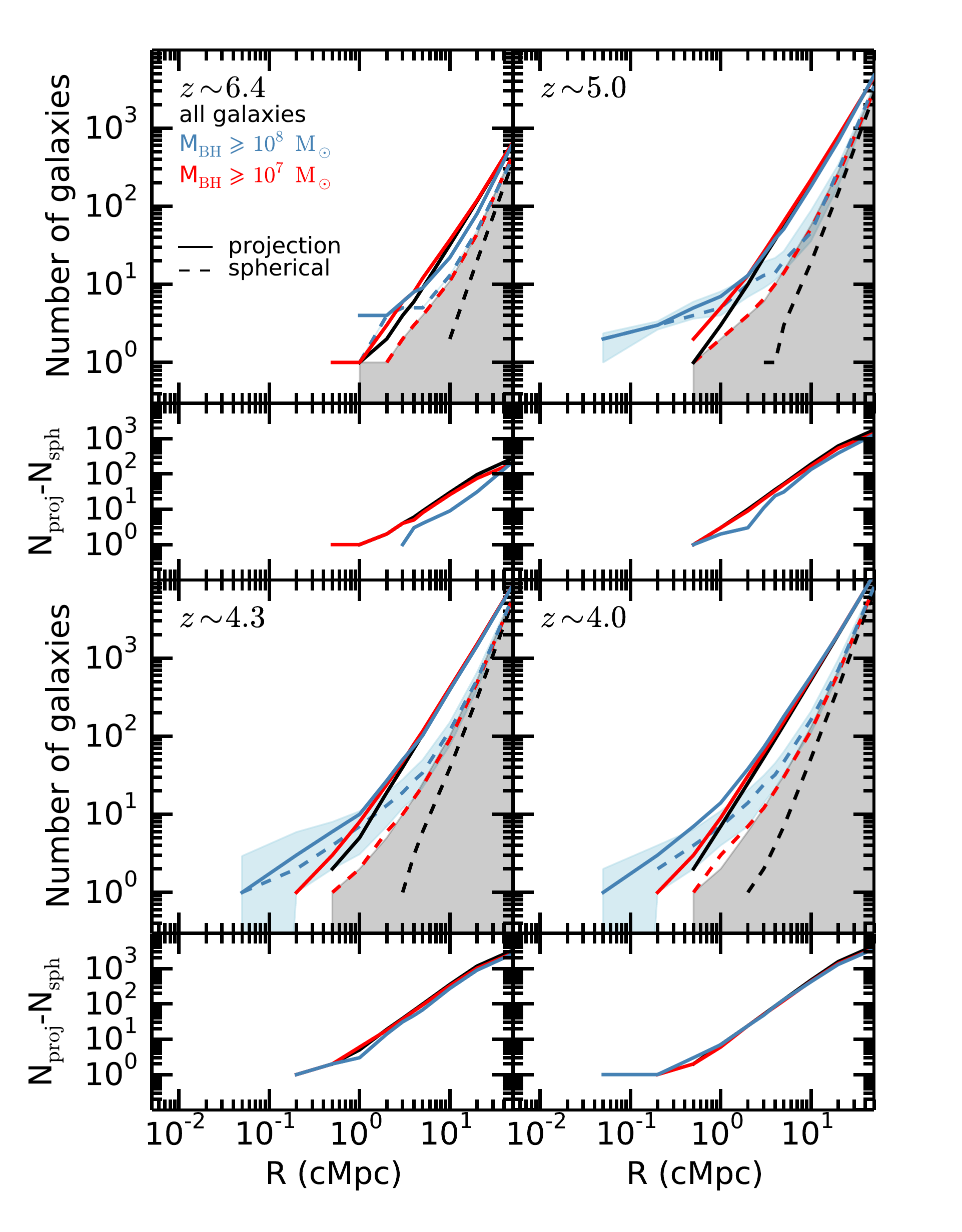}
\caption{Galaxy number counts around BHs in the large-scale cosmological simulation Horizon-AGN for two distinct methods to compute distances, either in sphere centered on galaxies (dashed lines), or as projections into the x-y plan (solid lines). Different quadrants are for different redshifts: $z=6.41$ (top left panel), $z=4.93$ (top right panel), $z=4.25$ (bottom left panel), and $z=3.98$ (bottom right panel). Here we consider BHs with $M_{\rm BH}\geqslant 10^7 \, \rm M_{\odot}$ (red lines), $M_{\rm BH}\geqslant 10^8 \, \rm M_{\odot}$ (blue lines). This does not include any selection on galaxy mass or luminosity, except the limit due to the resolution of the simulation. The blue shaded region represents the standard deviation $\pm 1\sigma$ for the 3D number count around $M_{\rm BH}\geqslant 10^8 \, \rm M_{\odot}$ BHs. Similar $\pm 1\sigma$ are found for the other cases, but are not shown here for clarity. Black lines show the median of average number counts for all galaxies, and the shaded black area the $\pm 1\sigma$ of the distribution. Finally, in the bottom panel of every quadrant, we also show the different galaxy number counts between the two methods, defined as $\rm N_{gal,projection}-N_{gal,spherical}$. The enhancement of counts around galaxies hosting massive BHs when using 3D separation distances, propagates when we use projected 2D distances, but with lower amplitude.}
\label{fig:nb_gal_radius_spherical_projected}
\end{figure*}

\subsection{Dark matter halo, galaxy and BH catalogues}
Dark matter haloes and sub-haloes are identified using AdaptaHOP halo finder~\citep{2004MNRAS.352..376A}, which uses an SPH-like kernel to compute densities at the location of each particle and partitions the ensemble of particles into sub-haloes based on saddle points in the density field. A total of 20 neighbours are used to compute the local density of each particles, and the density threshold is fixed to 178 times the average total matter density. The force softening (minimum size below which substructures are considered irrelevant) is $\sim 2\,\rm{kpc}$. Only dark mater haloes with more than 50 dark matter particles are considered.

Galaxies are identified following the same scheme and parameters, except that the stellar particle distribution is used instead of the dark matter one. We only consider galaxies with at least 50 stellar particles, which corresponds to $M_{\rm gal}\gtrsim 2\times 10^{8}\, \rm M_{\odot}$. The galaxy stellar mass functions at different redshift are presented in \citet{2017MNRAS.467.4739K}, where a good agreement is found with observations \citep{2012ApJ...752...66L,2015ApJ...810...73C,2016ApJ...825....5S} for $z\geqslant 4.0$, i.e. the redshift range that we are interested in. We note that the number of low-mass galaxies with $\log_{10} M_{\rm gal}\leqslant 9.5$ is somewhat overestimated in the simulation at $z=4$ and below.

Finally, we use the BH catalogues from \citet{2016MNRAS.460.2979V}, which studies the BH population of the simulation Horizon-AGN and provides a  comparison with observations (e.g. redshift evolution of the BH mass density, the BH mass function). See \citet{2016MNRAS.460.2979V} for a detailed descriptions of how the catalogue is created. In summary, a BH is assigned to a halo+galaxy structure if it lies within 10 per cent of the virial radius of the dark matter halo, and within twice the effective radius of the most massive galaxy within the halo. If more than one BH meets the criteria, the most massive BH is assigned as the central BH. 

In the following we focus our analysis on four high-redshift snapshots of the simulation, i.e. $z=6.4,5,4.3,4$, which span a similar redshift range as the observed high-redshift quasars described in Section 2.


\section{Galaxy number counts: methods}
In this section we describe the discrepancies between the 3D spherical number counts of galaxies, and the 2D number counts projected along the line-of-sight.

\subsection{3D spherical galaxy number counts}
We first evaluate the baseline against which to compare the counts around quasars and assess whether they are sitting in overdense regions. 
A first approach is to use the theoretical number of sources $N$ expected for galaxies randomly distributed in space.  
This can be defined by $N(R)=nV(R)$, with $n$ the number density of galaxies ($\rm cMpc^{-3}$) and $V$ the spherical volume of radius $R$ (cMpc). This represents the number of objects around a given galaxy if galaxies would not cluster. The {\it non-clustering} functions  for different redshifts are shown as dashed lines in Appendix~\ref{fig:reference_noclus}.

A second approach is to compare to the {\it average 3D number count}, defined as the median of the distribution of number counts around all the galaxies present in the simulation. Separation distances between galaxies are here computed as 3D distances, with spherical radii ranging from 0.05 cMpc to 50 cMpc. The simulation box being periodic, we correct galaxy positions accordingly for objects at the edges of the box. For simplicity, we do not account for the peculiar velocity of galaxies when we project their positions.
We show the {\it average 3D number count} as black dashed line in Fig.~\ref{fig:nb_gal_radius_spherical_projected} for different redshifts, the blue and black shaded areas represent the standard deviation $\pm \sigma$ of the distributions (only for two of the distributions for clarity). 
As galaxies tend to be clustered, the slopes of the {\it average 3D number count} functions are slightly shallower, and normalizations higher, than those of {\it non-clustering} functions (shown in Fig.~\ref{fig:reference_noclus}) due to the presence of an enhancement of galaxies at small scales (small $R$).

Following the same procedure, we compute the 3D number counts of galaxies around massive BHs, and show the median of the distribution in Fig.~\ref{fig:nb_gal_radius_spherical_projected} as colored dashed lines.
We use two BH mass ranges: $M_{\rm BH}\geqslant10^7 \, \rm M_{\odot}$ (red lines), and $M_{\rm BH}\geqslant10^8 \, \rm M_{\odot}$ (blue line). 
There are 81 $M_{\rm BH}\geqslant10^7 \, \rm M_{\odot}$ BHs at $z\sim6.4$, 934 at $z\sim 5$, 2019 at $z\sim4.3$, and 2737 at $z\sim4$. There is 1 $M_{\rm BH}\geqslant10^8 \, \rm M_{\odot}$ BH at $z\sim6.4$, 5 at $z\sim 5$, 33 at $z\sim4.3$, and 49 at $z\sim4$.
For clarity we only show the standard deviation $\pm \sigma$ (blue shaded area) of the distribution of the most massive BHs, but we find similar deviation for the $M_{\rm BH}\geqslant10^7 \, \rm M_{\odot}$ BH sample.
We find that, statistically, massive BHs are embedded in overdense regions, i.e. regions with an enhancement of the number count of galaxies compared to the {\it average 3D number counts} averaged over all galaxies (black dashed lines). When comparing the curves for the different BH mass ranges (red and blue dashed lines), we also note that on average the number counts around more massive BHs are even higher, and this for all redshifts.  


The difference in the 3D average galaxy number count and the number count around massive BH host galaxies is significant for spherical radius of $\rm R \lesssim 10\, \rm cMpc$. However, the enhancement of galaxy counts in the field of view of massive BHs becomes less statistically significant when we use projected 2D number counts. Projected 2D number counts are shown as solid lines in Fig.~\ref{fig:nb_gal_radius_spherical_projected} and explained in the following.\\

\subsection{Projected 2D galaxy number counts}
Computing the number of galaxies around a given galaxy in cosmological simulations is straightforward because the $x, y, z$ coordinates of galaxies are known. Observationally, the line-of-sight distance component is affected by redshift uncertainties. Here we aim at comparing results from the {\it 3D galaxy number counts} where we simply compute the spherical separation distance between galaxies to those from the {\it 2D galaxy number counts} where we use projected separation distance between galaxies, in better agreement with observations of high-redshift quasars.

We count the galaxies that are located around galaxies hosting a BH, in the projected plane $x-y$, for several circular radii ranging from 0.05 to $50\, \rm cMpc$. The z-axis corresponds to the line-of-sight, typically $\Delta z \sim~1$ for studies that use LBGs which corresponds to a line-of-sight distance $\Delta d \sim 450\, \rm{cMpc}$ 
at redshift $z \sim~6$ (the exact value depends on the cosmology). The constraint on $\Delta z$ is better when using LAEs, with $\Delta z \sim~0.1$ and $\Delta d\sim~ 44\, \rm{cMpc}$ at $z \sim~6$. 
The side length of Horizon-AGN is $142\, \rm{cMpc}$. In our analysis we use $dz=50\, \rm{cMpc}$. With this choice, we will underestimate the number of projected objects of the quasars field of view compared to the observations of LBGs when their redshift is not well constrained. At the end of the next section, we discuss another choice of $dz$ to understand the impact of redshift uncertainty in the galaxy number counts.

Again, we first evaluate references for the galaxy number counts. We first derive the {\it projected non-clustering galaxy number counts}, which assumes that galaxies are randomly distributed in the simulation: $N(R)=n_{\rm proj}V_{\rm proj}(R)=n_{sph}\times dz \times S_{\rm proj}(R)$, with $n$ the mean number density of galaxies in the whole simulation volume per $\rm cMpc^{-3}$, $dz$ the length of the line of sight redshift uncertainty (here $dz = 50 \rm \, cMpc$), $S_{\rm proj}$ the projected surface between two galaxies of projected separation $R$ (cMpc). 
These {\it projected non-clustering} functions are shown as solid lines in Appendix~\ref{fig:reference_noclus}, their normalization and slope is slightly higher than the 3D functions due to the projection effect.
A second approach is to compare to the {\it average 2D number counts}, whose normalization is higher than the previous because galaxies tend to be clustered. To do so, we compute the number of galaxies in projected circles centered on each galaxy center, and of increasing radius $R$ from $0.05$ to $50\, \rm{cMpc}$, for all galaxies within the simulation box, hosting or not a BH, and we take the median of this distribution. 
We show the {\it average 2D counts} as solid black lines in Fig.~\ref{fig:nb_gal_radius_spherical_projected} for different redshifts. In the following figures we use the second approach as a reference.  

As for the 3D procedure, the median of the distributions representing the 2D number of galaxies around massive BHs in the simulation are higher than the {\it average number counts} derived from the simulation on scales $\lesssim 10\, \rm{cMpc}$, with counts being higher the higher is the BH mass threshold considered, but the difference is of order a few galaxies and it is statistically undistinguishable from the average case. 
While the deviation of the 3D number counts in the field of view of massive BHs compared to the average 3D number counts was statistically significant, we find that the deviation of the 2D number counts around massive BHs is not significant anymore and is for most radii $R$ within the variance of the average 2D number count functions.

\begin{figure*}
\centering
\includegraphics[scale=0.81]{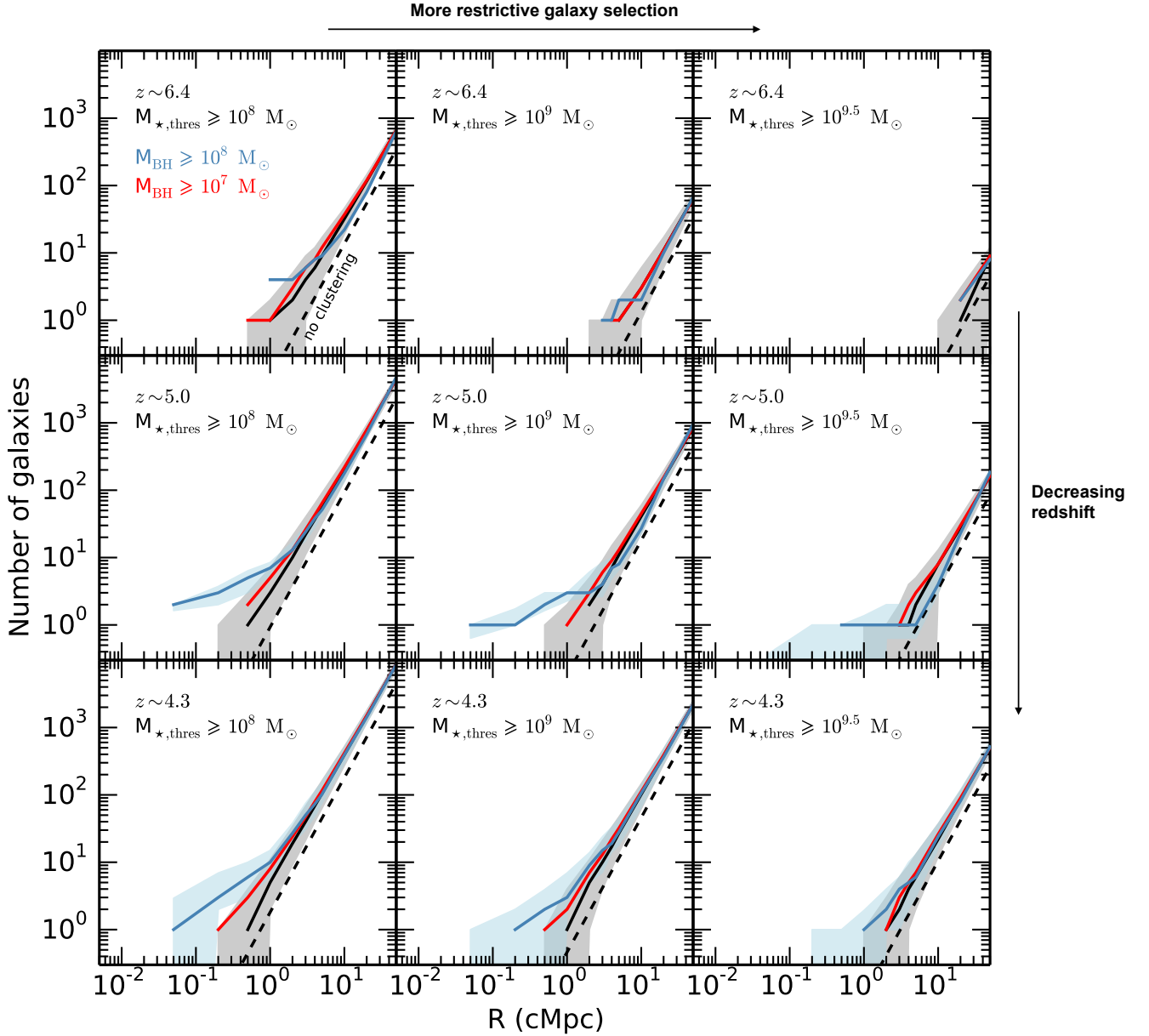}
\caption{Counts of galaxies around BHs in Horizon-AGN at $z=6.41$ (top row), $z=4.93$ (middle row), and $z=4.25$ (bottom row), for 3 different stellar mass thresholds to select surrounding galaxies $M_{\rm \star, thres}=10^{8}\, \rm M_{\odot}$ (left column), $M_{\rm \star, thres}=10^{9}\, \rm M_{\odot}$ (middle column),  and $M_{\rm \star, thres}=10^{9.5}\, \rm M_{\odot}$ (right column). Here we use a projected length of 50 cMpc. Black solid lines represent the counts averaged over all the galaxies present in the simulation box, we use these lines as reference of regions with average galaxy number density. Similarly we also show the theoretical number counts of galaxies in a non-clustered field as black dashed lines. Here we consider BHs with $M_{\rm BH}\geqslant 10^7 \, \rm M_{\odot}$ (red line), $M_{\rm BH}\geqslant 10^8 \, \rm M_{\odot}$ (blue line). 
The blue shaded region represents the standard deviation $\pm 1\sigma$ for the particular case of $M_{\rm BH}\geqslant 10^8 \, \rm M_{\odot}$. Similar $\pm 1\sigma$ are found for the other cases, but are not shown here for clarity.}
\label{fig:pict_1}
\end{figure*}

\section{Galaxy number counts: results}
In this section we first test the visibility of the enhancement of the galactic 2D projected number counts around massive BHs. 
While in simulations we can count the {\it true} number of galaxies around massive BHs, observations rely on the instruments used for the measure, i.e. a given capability, and spectral band(s). In order to keep our results as general as possible, in the following we do not select galaxies as a function of their luminosity or magnitude in a given spectral band, but rather choose to use the stellar mass of galaxies.
In observations, several methods have been used to select galaxies in the environment of massive BH host galaxies. We summarize these methods in Table~\ref{obs_methods}. At high redshift it is convenient to select galaxies based on their UV emission from young stellar populations, but the uncertain dust content of these galaxies \citep{2017PASA...34...31V,2017Natur.545..457D} leads to large uncertainties in the rest-frame UV luminosity/magnitude computed from simulated galaxies. Selections can also be based on colors, as used in a large number of observational studies (although all using different color selection criteria). Colors are more well constrained in observations than galaxy stellar mass, but again are difficult to derive accurately from simulations \citep[see][for more discussion on the uncertainties of computing colors from simulations]{2018MNRAS.475..624N}. Given these large uncertainties and the very diverse magnitude/color selections that have been employed in observations so far, we prefer to simply rely on galaxy stellar mass to select galaxies.


We apply three stellar mass thresholds to select galaxies in the field of view of the massive BHs, those are chosen to fairly represent what we be achievable with upcoming space missions: 
\begin{itemize}
\item $M_{\star}\geqslant 10^{8}\, \rm M_{\odot}$, which is similar to the capacity of JWST\footnote{In theory JWST should even be able to detect $M_{\star}\geqslant 10^{7}\, \rm M_{\odot}$, but we are here limited by the resolution of the simulation.};
\item $M_{\star}\geqslant 10^{9}\, \rm M_{\odot}$ to mimic the capacity of the upcoming mission WFIRST;
\item $M_{\star}\geqslant 10^{9.5}\, \rm M_{\odot}$ to mimic the capacity of the mission Euclid.
\end{itemize}

We have selected these thresholds based on \citet{2018ApJS..236...33W,2017jwst.prop.1345F} for JWST, \citet{2015arXiv150303757S} for WFIRST, \citet{2016SPIE.9904E..0OR} for Euclid, and the theoretical predictions for these missions derived in Yung et al. (in prep).
The percentage of galaxies with a stellar mass equal or higher than these thresholds in the simulation Horizon-AGN, is respectively: $99.8\%$, $9.7\%$ and $1.3\%$, at redshift $z\sim6.4$. We find similar fractions for the following snapshots: $99.9, 19.4, 3.7\, \%$ at $z\sim5$, and $99.9, 25.5, 5.8\%$ at $z\sim4.3$.

\begin{figure*}
\includegraphics[scale=0.62]{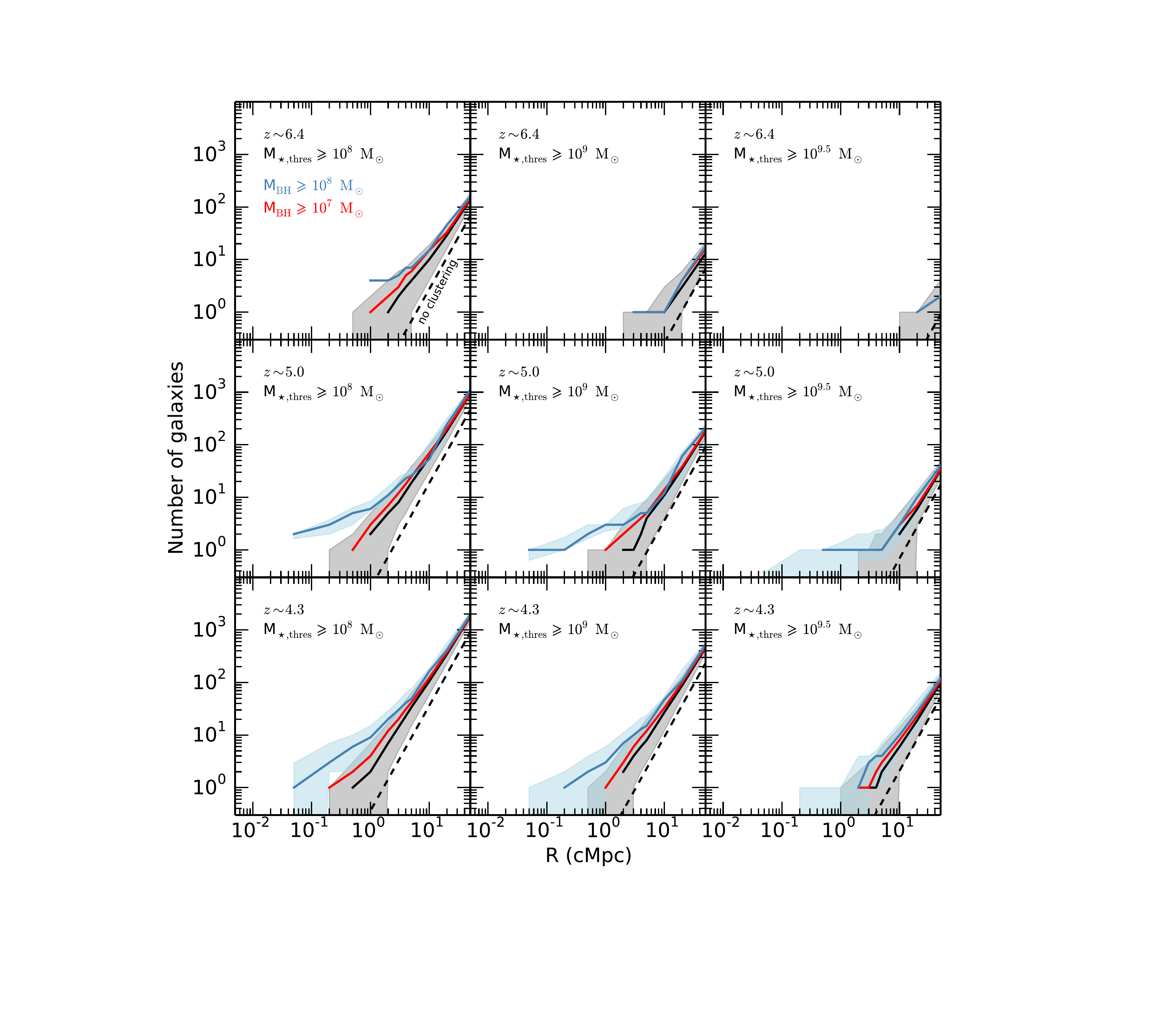}
\caption{Counts of galaxies around BHs in Horizon-AGN at $z=6.41$ (top row), $z=4.93$ (middle row), and $z=4.25$ (bottom row). Same as Fig~\ref{fig:pict_1} but with a project length of 10 cMpc instead of 50 cMpc. The enhancement of the galaxy number counts of the most massive BHs is higher for the two lower stellar mass thresholds.}
\label{fig:smaller_length}
\end{figure*}

\begin{table*}
\caption{Percentage $P$ of massive BHs ($M_{\rm BH}\geqslant 10^{8}\, \rm M_{\odot}$) having their nearest galaxy neighbor enclosed in a characteristic shell, which is defined with the projected separation distances $R$ = 0.05, 0.2, 0.5, 1, 2, 3, 4, and 5 cMpc. We use the three stellar mass thresholds to select galaxies: $M_{\star, \rm thres}=10^{8}, 10^{9}, 10^{9.5}$. For comparison we also report the percentage $P_{\rm ref}$ for galaxies of same mass, but hosting lower mass BHs ($M_{\rm BH}< 10^{8}\, \rm M_{\odot}$).} 
\begin{center}
\begin{tabular}{ccccccccccccccc}
\hline

\multicolumn{1}{c}{z} & \multicolumn{1}{c}{$R$ (cMpc)} && \multicolumn{2}{c}{$M_{\rm \star,thres}=10^{8}\, \rm M_{\odot}$} && \multicolumn{2}{c}{$M_{\rm \star, thres}=10^{9}\, \rm M_{\odot}$} && \multicolumn{2}{c}{$M_{\rm \star, thres}=10^{9.5}\, \rm L_{\odot}$}\\
& && $P$ &  $P_{\rm ref}$ && $P$ &  $P_{\rm ref}$ && $P$ &  $P_{\rm ref}$ \\
\hline
\hline
                                 & 0 - 0.05 && 100 & 100 && 80 & 80 && 20 & 20 \\
                                 & 0.05 - 0.2 && 0 & 0 && 20 & 20           && 20 & 20\\
$z\sim 5.0$ (5 BHs) & 0.2 - 0.5 && 0 & 0 && 0 & 0               && 20  & 20\\
                                 & 0.5 - 1    && 0 & 0 && 0 & 0               && 0   & 20\\
                                 & 1 - 2     && 0 & 0 && 0 & 0              && 0   & 20\\
\hline
                                 & 0 - 0.05 && 73 & 40 && 42 & 21    && 12 & 9\\
                                 & 0.05 - 0.2 && 21 & 36 && 24 & 29      && 15 & 12\\
$z\sim4.3$ (33 BHs) & 0.2 - 0.5 && 6 & 18 && 12 & 23        && 15 & 20\\
                                 & 0.5 - 1    && 0 & 6 && 18 & 18          && 9 & 18\\
                                 & 1 - 2    && 0 & 1 && 3 & 7              && 39 & 26\\
\hline

\end {tabular}
\end{center}
\end{table*}

\begin{table*}
\caption{Percentage of low density environments of massive BHs ($M_{\rm BH}\geqslant 10^{8}\rm M_{\odot}$), i.e. environments that have a galaxy number counts $\rm N_{gal}$ smaller than the average number counts $\rm N_{ref}$ ( $\rm N_{ref}$ is shown as black lines in Fig.~\ref{fig:nb_gal_radius_most_clustered_lumcut}) within a 2D projected separation distance of 1 cMpc and 2 cMpc. We include the percentages for redshift $z\sim 5$, and $z\sim 4.3$, and for the two selection limits $M_{\star}=10^{8}\, \rm M_{\odot}$ and $M_{\star}=10^{9.5}\, \rm M_{\odot}$.}
\begin{center}
\begin{tabular}{ccccc}
\hline

& \multicolumn{2}{c}{$z\sim5$} & \multicolumn{2}{c}{$z\sim 4.3$} \\
& $M_{\star}=10^{8}\, \rm M_{\odot}$ & $M_{\star}=10^{9.5}\, \rm M_{\odot}$ & $M_{\star}=10^{8}\, \rm M_{\odot}$ & $M_{\star}=10^{9.5}\, \rm M_{\odot}$ \\

\hline
\hline
$\rm N_{gal}\, \leqslant\, N_{ref}$ at 1 cMpc     & $0 \%$ & $40 \%$ & $12\%$ & $48\%$\\
$\rm N_{gal} \, \leqslant \, N_{ref}$ at 2 cMpc      & $0 \%$ & $20\%$ & $21\%$ & $39\%$ \\

\hline
\end {tabular}
\end{center}
\label{table_percentage_lowdens}
\end{table*}

\subsection{Number counts around massive BH host galaxies}

Fig.~\ref{fig:pict_1} shows the number counts of galaxies around massive BHs as a function of 2D projected separation distance R (cMpc), ranging from 0.05 to 50 cMpc. We use a projected length of $dz=50\, \rm cMpc$. Top panels present our results for $z\sim 6.4$, middle panels for $z\sim 5$, and bottom panels for $z\sim4.3$. We increase the stellar mass threshold from left to right panels, as labelled in the figure.
We find that the normalization of the counts evolve both with time and with galaxy stellar mass threshold selection. With decreasing redshift from $z\sim 6$ to $z\sim 4$ (bottom panels), more galaxies are present in the simulation, and therefore the normalization of the counts increases.
Similarly, the number of massive galaxies in the field of view decreases when we increase the selection threshold.

Galaxy number counts at small scales are higher in the environment of the most massive BHs (blue lines), and the lower the galaxy stellar mass cut the stronger the contrast between counts for different BH mass cuts and an average density region. Therefore, the more massive BHs are statistically embedded in overdense regions. The distribution of the counts for $M_{\rm BH}\geqslant 10^{8}\, \rm M_{\odot}$ presents an overdensity down to  $\rm R\sim 10^{-1}\, cMpc$. This means that at least few regions in the simulation box show very dense environments, and those are primarily due to satellite galaxies, i.e. galaxies that are embedded in the same dark matter halo as the central galaxies hosting the massive BHs that we are looking at. However, differences among the various cases are within the variance and almost disappear when considering a high galaxy stellar mass cut of $M_{\star}\geqslant 10^{9.5}\, \rm M{\odot}$. If the trend with mass observed here continued at higher BH mass (i.e. quasars), a statistically significant number of deep observations of the fields of bright quasars should be able to see an enhancement on scales $\lesssim 10\, \rm cMpc$. This should be achievable with JWST and WFIRST.

Here we have used a projected length of 50 cMpc, which correspond to the redshift uncertainty that can be obtained observationally with LAEs ($\Delta z \sim 0.1$). While photometric redshifts at high redshift ($z\sim 5$) are less accurate, i.e. in the range $\Delta z \sim 0.1-0.3$ (private communication with Jeffrey Newman, for future missions such as WFIRST), redshift estimate with grisms are more precise, in the range $\Delta z < 0.05$. To understand how the number counts of galaxies is affected by redshift uncertainty, we show the analog of Fig.~\ref{fig:pict_1} in Fig.~\ref{fig:smaller_length}, but using a projected length of 10 cMpc, i.e mimicking a much better accuracy of the redshift estimate. At redshift $z\sim6$, a redshift uncertainty of $\Delta z <0.05$ (grisms) would correspond to $\Delta d < 20 \, \rm cMpc$.
With this better accuracy of redshift estimate, we find that the enhancement of the galaxy number counts is higher (the blue lines compared to the black lines), and visible to larger scales, particularly for the lower stellar mass thresholds $M_{\star}\geqslant 10^{8}\, \rm M_{\odot}$ and $M_{\star}\geqslant 10^{9}\, \rm M_{\odot}$. For those thresholds the difference between the number counts around BHs of $M_{\rm BH}\geqslant 10^{7}\, \rm M_{\odot}$ and $M_{\rm BH}\geqslant 10^{8}\, \rm M_{\odot}$ is also larger.

We also want to emphasize here that observations probe the environment of bright quasars. 
Because AGN luminosity is subject to high variability in time, we selected our samples on BH mass, considering only the most massive BHs of $M_{\rm BH}\geqslant 10^{7}\, \rm M_{\odot}$ in the present paper. 
To investigate a possible bias of BH activity on the number counts of galaxies in the BH environments, we split our BH sample into an AGN sample and a more quiescent BH sample.
AGN are defined here as having a bolometric luminosity of $\log_{10} L_{\rm bol}\geqslant 44$ at the time of the current snapshot or at some point within the last 25 Myr. 
We do not find significant differences within these two samples: accreting BHs do not have a particularly higher number counts of galaxies in their field of view, on average. Our conclusions are the same if we consider a BH as being an AGN in the current snapshot or as having experienced an AGN phase in the past (within the last 25 Myr).

\subsection{Closest galaxy companion}
As explained above, we find that the difference between the average number counts and the number counts in the environment of the most massive BHs is only significant at scale below few cMpc. In Table 1, we quantify the probability $P$ for a galaxy hosting a $\geqslant 10^{8}\, \rm M_{\odot}$ BH to have a closest galaxy companion enclosed in shells of characteristic projected separation radii of $R=0 - 0.05, 0.05 - 0.2, 0.2 - 0.5, 0.5 - 1, 1 - 2\, \rm cMpc$. 
For comparison we provide the probability of having a close companion galaxy in a given shell $P_{\rm ref}$, for galaxies of same mass but hosting a lower-mass BHs of $< 10^{8}\, \rm M_{\odot}$. We compute these probabilities assuming the three different stellar mass thresholds.
At $z\sim 5$, we find that the probabilities $P$ and $P_{\rm ref}$ of having a close galaxy companion are very similar. These results are affected by the very low number of $M_{\rm BH}\geqslant 10^{8}\, \rm M_{\odot}$ BHs at this redshift.  
At lower redshift $z\sim 4.3$ for which we have better statistics, we find that galaxies hosting massive BH have a higher probability $P$ of having a close galaxy companion (e.g., at $R=0.05\, \rm cMpc$) than galaxies with lower mass BHs or no BHs. The discrepancy is slightly washed out when using more restrictive galaxy selection, such as $M_{\star, \rm thres}=10^{9.5}\, \rm M_{\odot}$.
Recent investigations of the environment of high-redshift quasars \citep{2017ApJ...836....8T,2017Natur.545..457D} indeed find close companions with separation distances of $\leqslant 0.1 \, \rm cMpc$ (i.e. $\leqslant 50 \, \rm pkpc$ in proper distances). We confirm that such close companions should be expected.

\begin{figure*}
\centering
\includegraphics[scale=0.56]{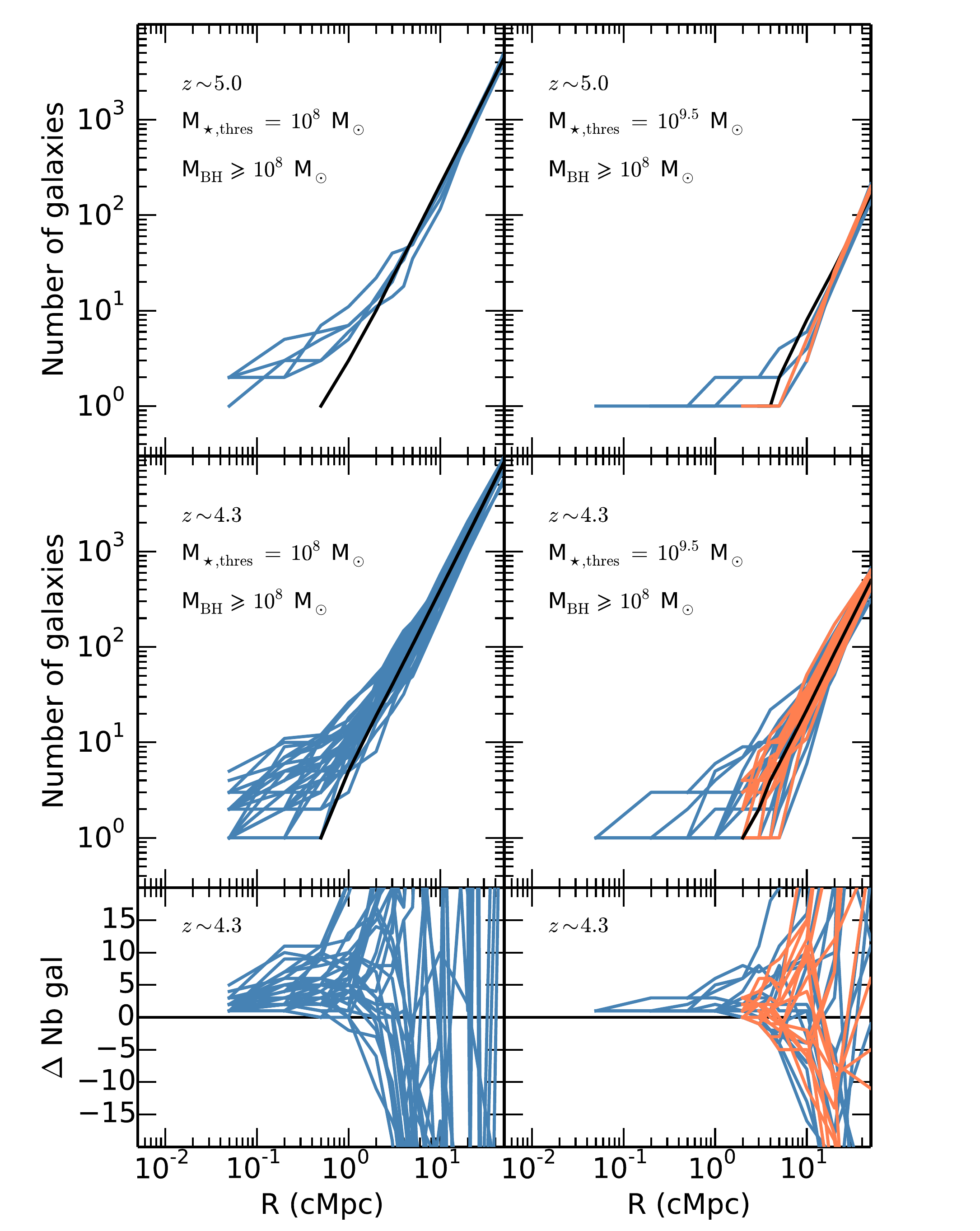}
\caption{Galaxy number counts around individual BHs of $M_{\rm BH}\geqslant 10^{8}\, \rm M_{\odot}$ with stellar mass cut of $M_{\rm \star, thres}=10^{8}\, \rm M_{\odot}$ (left panels) and $M_{\rm \star, thres}=10^{9.5}\, \rm M_{\odot}$ (right panels) in blue solid lines, at $z\sim 5$ (top panels), and $z\sim 4.3$ (middle panels). We highlight with orange solid lines the number counts for which there is no galaxy within 1 cMpc, i.e. $N(R\leqslant 1\, \rm cMpc)=0$. In each panel, the black line represent the corresponding average galaxy number counts. The bottom panels show the difference number counts between the environment of black holes and the average galaxy number counts (black lines in the top panels).}
\label{fig:nb_gal_radius_most_clustered_lumcut}
\end{figure*}

In the simulation we also find a few BHs at $z\sim 4.3$ that have a close galaxy companion and a not particular enhanced number counts of galaxies at larger projected distances from the BHs. 
We define these as the systems having $P>P_{\rm ref}$ for the shell of separation distances $0-0.2\, \rm cMpc$ but $N_{\rm gal}\sim N_{\rm ref}$ at $1\, \rm cMpc$. 
At $z\sim 5$, the percentage of such systems is $0, \,20, \,40\, \%$, and at $z\sim 4.3$ there are $6, \,18,\, 21\,\%$ for $M_{\star, \rm thres}=10^{8},\, 10^{9}, \,10^{9.5}\, \rm M_{\odot}$ respectively.
One of the BHs with $M_{\rm BH}\geqslant 10^{8}\, \rm M_{\odot}$ has a close galaxy neighbor at $R=0.2\, \rm cMpc$ and the closest next galaxies are observed at projected distances of $R\sim 1\, \rm cMpc$. Similarly, we also identify a BH of the same mass with two nearby galaxies at $R\sim 0.2\, \rm cMpc$, and no other galaxies before $R\sim 1 \, \rm cMpc$. These two BHs have 5 and 6 galaxies, respectively, located at distances of $R\sim 1\rm cMpc$, which is in agreement with the average number of galaxies to be expected at such distance (5 expected galaxies on average, black lines in Fig.~\ref{fig:nb_gal_radius_most_clustered_lumcut}). 
These examples assumed a low stellar mass threshold of $M_{\star, \rm thres}=10^{8}\, \rm M_{\odot}$ that enables the detection of low-mass galaxies in the BH fields of view. For higher stellar mass thresholds (e.g., $M_{\star, \rm thres}=10^{9.5}\, \rm M_{\odot}$), some nearby low mass galaxies can be missed, which makes the cases described above more common. We detail this in the next section. 

The presence of close companions may be related to merging activity. One could ask whether the environment had been overdense in the past, and whether the companion galaxies have now merged with the host of the BH. Regarding past merging activity, for the first of the examples above, the environment was not particularly denser in the past between $z\sim6.4-4.3$, and the closest neighbor always located at $\geqslant 1\, \rm cMpc$ away. The BH and its host galaxy have grown by less than one order of magnitude during this time.
However, the past environment of the second BH was denser with a very close companion at $R=0.05\, \rm cMpc$ and one other galaxy within a projected distance of $0.2\, \rm cMpc$ at $z\sim 5$. The closest galaxy merged with the BH host galaxy, and both the BH and its host galaxy have grown by more than one order of magnitude in the redshift range $z\sim6.4-4.3$.  

\subsection{The diverse environments of high-redshift massive BHs}
Given the limited number of deep observations of bright quasar field the relevant approach is to examine the diversity of number counts around individual BH hosts. We show galaxy number counts around individual $\geqslant 10^{8}\, \rm M_{\odot}$ BH host galaxies as solid blue lines in Fig.~\ref{fig:nb_gal_radius_most_clustered_lumcut}. Top panels represent $z\sim 5$, middle panels $z\sim 4.3$, and we show the difference number counts between the environment of massive BHs and the average number counts (shown in solid black lines) for $z\sim 4.3$ in the bottom panels. Left panels shows results when we use a stellar mass cut of $M_{\rm \star, thres}=10^{8}\, \rm M_{\odot}$, and right panels use $M_{\rm \star, thres}=10^{9.5}\, \rm M_{\odot}$. 
We highlight in orange the environments that do not have any galaxy count for $R\leqslant \, \rm 1cMpc$.
We do not find any case like that at $z\sim 5$ or $z\sim 4.3$ when we consider a low stellar mass cut of $M_{\rm \star, thres}=10^{8}\, \rm M_{\odot}$. However, when we increase the stellar mass threshold to $M_{\rm \star, thres}=10^{9.5}\, M_{\odot}$, we find that $40\%$ of BH fields of view have a null number count of galaxies within a 2D separation distance of 1 cMpc (i.e. $N(R\leqslant 1\, \rm cMpc)=0$) at $z\sim 5$, and $48\%$ at $z\sim 4.3$.
We report the percentage of the low density environments around massive BHs ($M_{\rm BH}\geqslant 10^{8}\, \rm M_{\odot}$), i.e. those that have galaxy number counts $\rm N_{gal}$ below the average expectation $\rm N_{ref}$\footnote{Since the reference is averaged over all the simulated galaxies in the box (hosting a BH or not), and that we study the number counts in the fields of the massive BHs, we do not expect to find the same number of fields below and above the reference.} in Table~\ref{table_percentage_lowdens} within separation distance of 1 cMpc or 2 cMpc. The reference $\rm N_{ref}$ corresponds to the median value of the number counts in the fields of view of all galaxies in the simulation, and is shown as black lines in Fig.~\ref{fig:nb_gal_radius_most_clustered_lumcut}. 

There are two important results here. First of all, we find that even if on average the fields around massive BHs contain more galaxies than control fields, it exits a diversity of environments when we study individual field of view. Some fields have a large number counts of galaxies, while other fields have a galaxy number counts very close to the average number counts around galaxies (hosting or not a BH).
Second, we find that the diversity of environments is exacerbated when the observational capacities only allow us to observe the most massive or brightest galaxies in the field of view. Indeed, the number of galaxies close ($R<1\, \rm cMpc$, orange lines) to the BH host galaxies decreases strongly when using a restrictive stellar mass cut ($M_{\rm \star, thres}=10^{9.5}\, \rm M_{\odot}$).

\begin{figure*}
\centering
\includegraphics[width=\columnwidth]{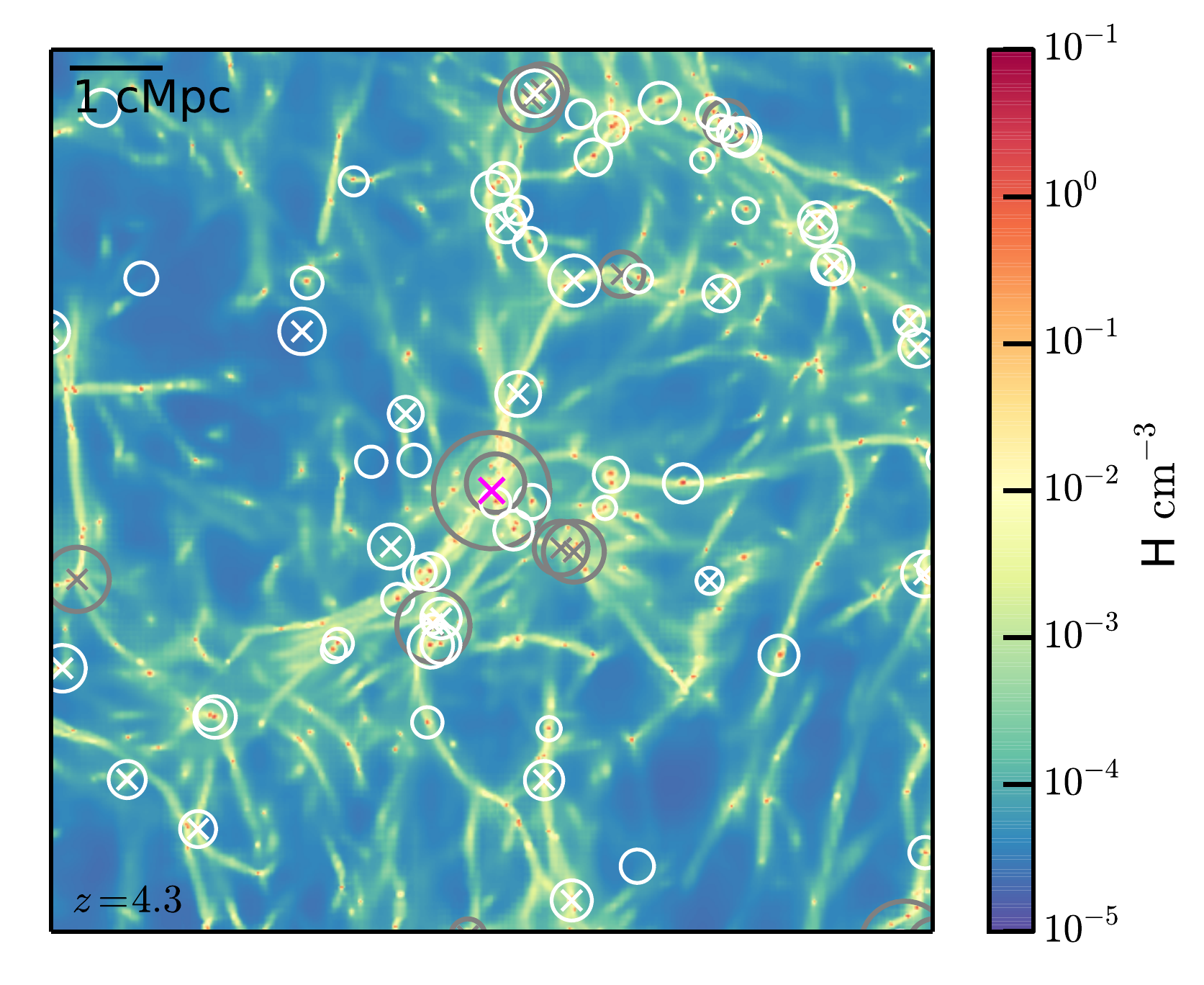}
\includegraphics[width=\columnwidth]{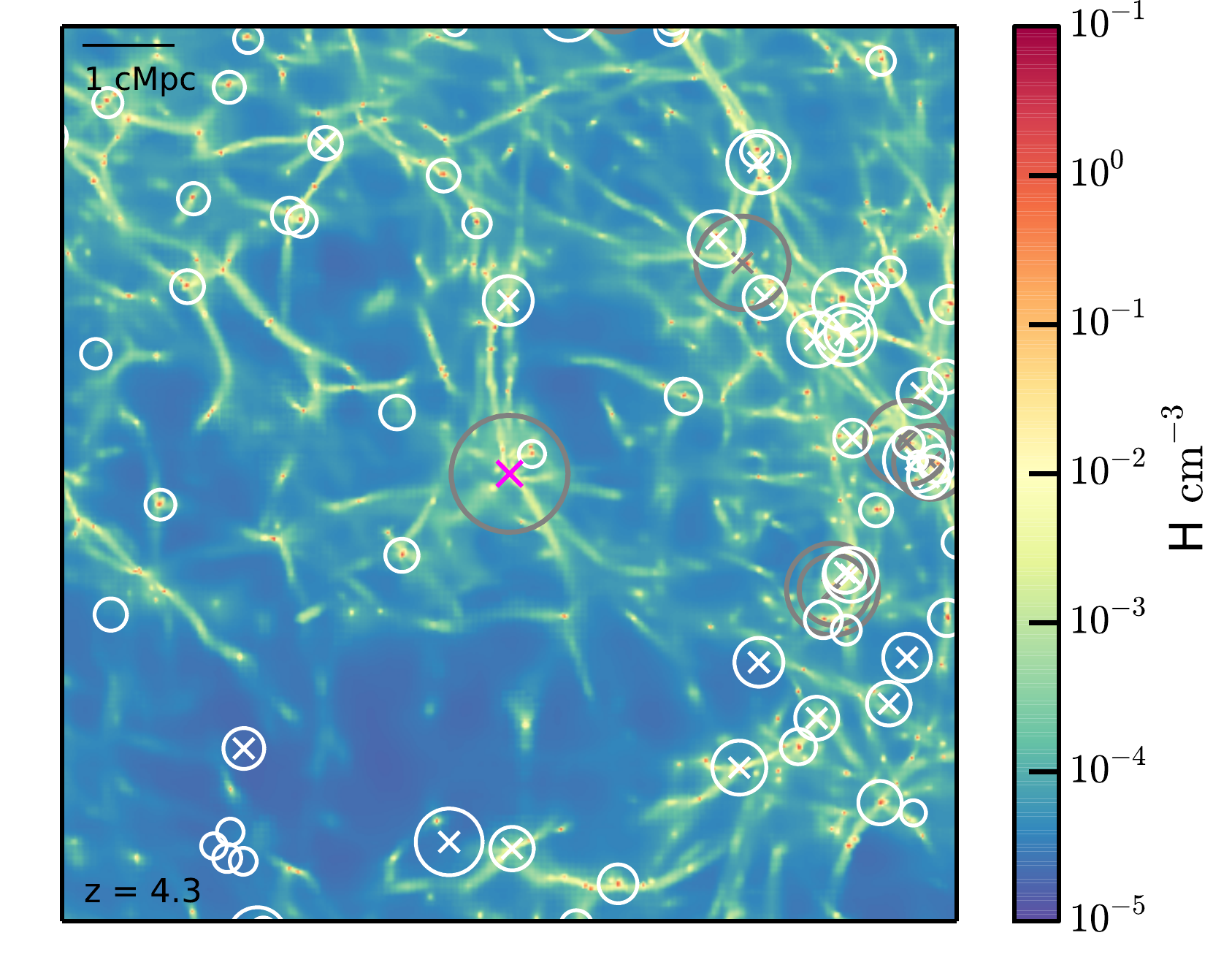}
\caption{{\it Left panel:} example of a massive BH (pink cross) embedded in a very dense region at $z=4.3$. 
{\it Right panel:} example of a massive BH embedded in an under-dense region, at the same redshift. The figures represent gas density maps in unit of $\rm H/cm^{3}$. Neighbouring galaxies are shown with white circles if $M_{\star }\geqslant 10^{8} \, \rm M_{\odot}$, and grey circles if $M_{\star}\geqslant 10^{9.5}\, \rm M_{\odot}$. 
The central grey circle in both figures indicate the host galaxy of the studied massive BH. 
The circle radii are equal to 10 times the virial radius of galaxies.
Galaxies hosting a central BH are indicated with a cross. Here, we have selected these two examples as having many or very few galaxies at a separation distance smaller than $2 \, \rm cMpc$, considering galaxies with $M_{\rm \star, thres}=10^{8}\, \rm M_{\odot}$. }\label{fig:individual_clus}
\end{figure*}

To highlight the diversity of environments that we find in Horizon-AGN, we provide two examples in Fig.~\ref{fig:individual_clus}. The left panel indicates a dense environment of a massive BH with $M_{\rm \star, thres}=10^{8}\, \rm M_{\odot}$, while the right panel represents an underdense region around another BH of the same mass. 
Here, we have selected these two examples as having many or very few galaxies at a separation distance smaller than $2 \, \rm cMpc$.
Galaxies in the field of view are shown with white circles if $M_{\star }\geqslant 10^{8} \, \rm M_{\odot}$, and grey circles if $M_{\star}\geqslant 10^{9.5}\, \rm M_{\odot}$. 
The central grey circle in both panels indicate the host galaxy of the studied massive BH. 
The circle radii are equal to 10 times the virial radius of galaxies.
In the right panel, the galaxies within a couple of cMpc away from the BH are almost all galaxies with $M_{\star}\leqslant 10^{9}\, \rm M_{\odot}$ (except the one at $\sim 2\, \rm cMpc$ right on top of the BH host galaxy). Therefore this region will appear almost entirely devoid of nearby galaxies, with probably a ring-like shape at $\sim 3.5 \, \rm cMpc$, with instruments observing galaxies of $M_{\star}\geqslant 10^{9}\, \rm M_{\odot}$, such as WFIRST or Euclid will do.

\begin{figure}
\centering
\includegraphics[scale=0.56]{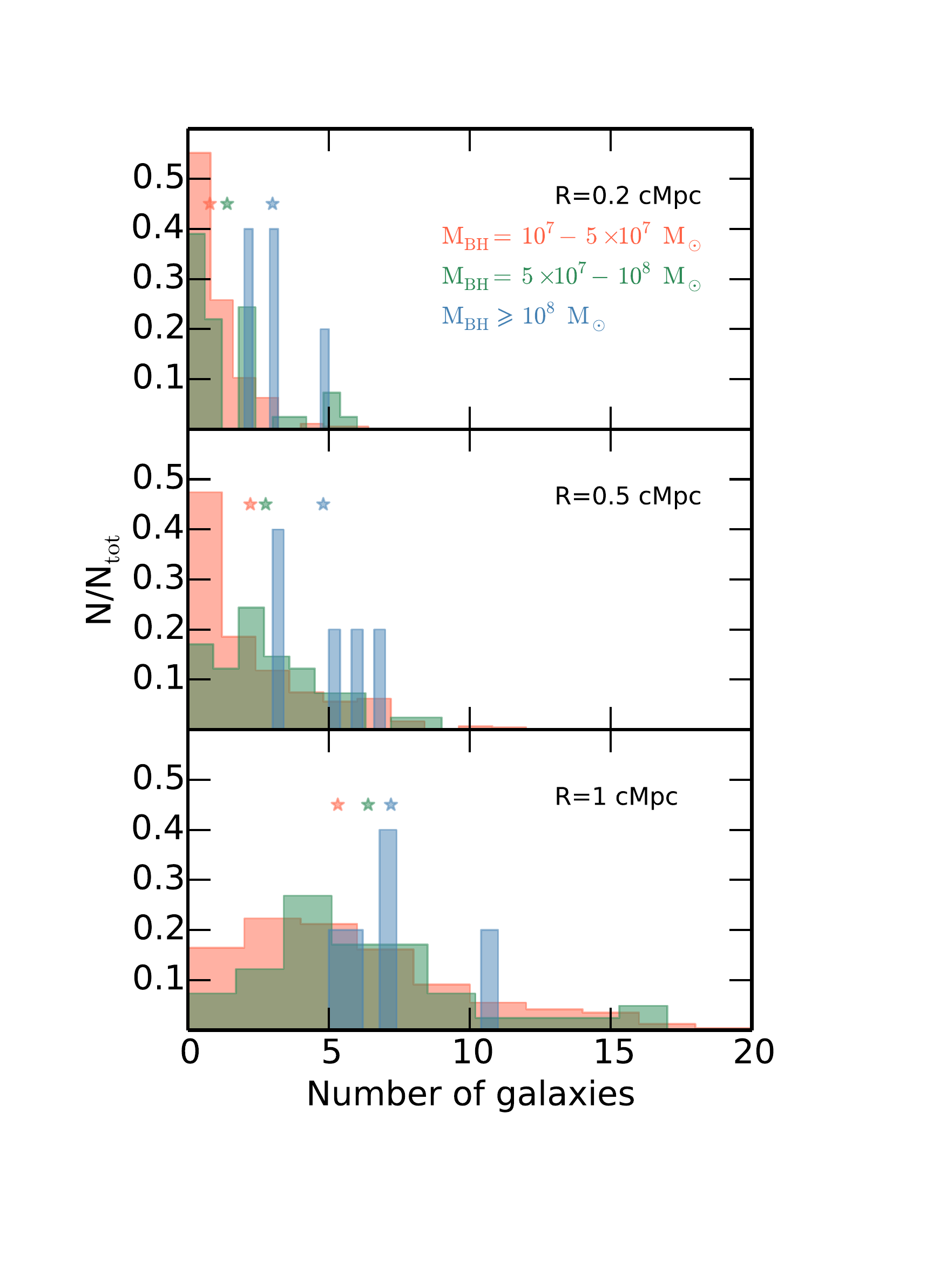}
\caption{Normalized distributions of the galaxy number counts at $z\sim5$ around massive BHs of $M_{\rm BH}=10^{7}\,- 5\times 10^{7} \rm \,  M_{\odot}$ in red, $M_{\rm BH}=5\times 10^{7} - 10^{8}\rm \, M_{\odot}$ in green, and $M_{\rm BH}\geqslant 10^{8}\, \rm M_{\odot}$ in blue. Here we use a stellar mass cut of $M_{\rm \star, thres}=10^{8}\, \rm M_{\odot}$. Galaxy number counts are computed within projected separation distances of $R=0.2\, \rm cMpc$ (top panel), $R=0.5\, \rm cMpc$ (middle panel), and $R=1\, \rm cMpc$ (bottom panel).
Star symbols show the mean of the distributions.}
\label{fig:histo}
\end{figure}

Finally, we also present in Fig.~\ref{fig:histo} the distributions of the galaxy number counts for different BH mass bins, i.e. $M_{\rm BH}=10^{7}-5\times 10^{7}\, \rm M_{\odot}$ in red, $M_{\rm BH}=5\times10^{7}- 10^{8}\, \rm M_{\odot}$ in green, and $M_{\rm BH}\geqslant10^{8}\, \rm M_{\odot}$ in blue. The galaxy number counts are summed within a given projected separation distance R: $\rm R=0.2 \, cMpc$ in the top panel, $\rm R=0.5 \, cMpc$ in the middle panel, and $\rm R=1 \, cMpc$ in the bottom panel. Here, we show the results for $z=5$ and for a selection limit of $M_{\star}=10^{8}\, \rm M_{\odot}$, but similar results are obtained for different redshift and stellar mass cuts. The mean values of the distributions are indicated with star symbols. Again, the figure demonstrates that on average more massive BHs live in regions with more neighborhood galaxies.  Only $\sim 20$ quasar fields of view have been observed so far. 
Selecting randomly 20 of these BH fields of view in the simulation would also provide us with diverse BH environments. This is true for all projected separation distances $R=0.2, 0.5, 1\, \rm cMpc$.

\begin{figure*}
\centering
\includegraphics[scale=0.53]{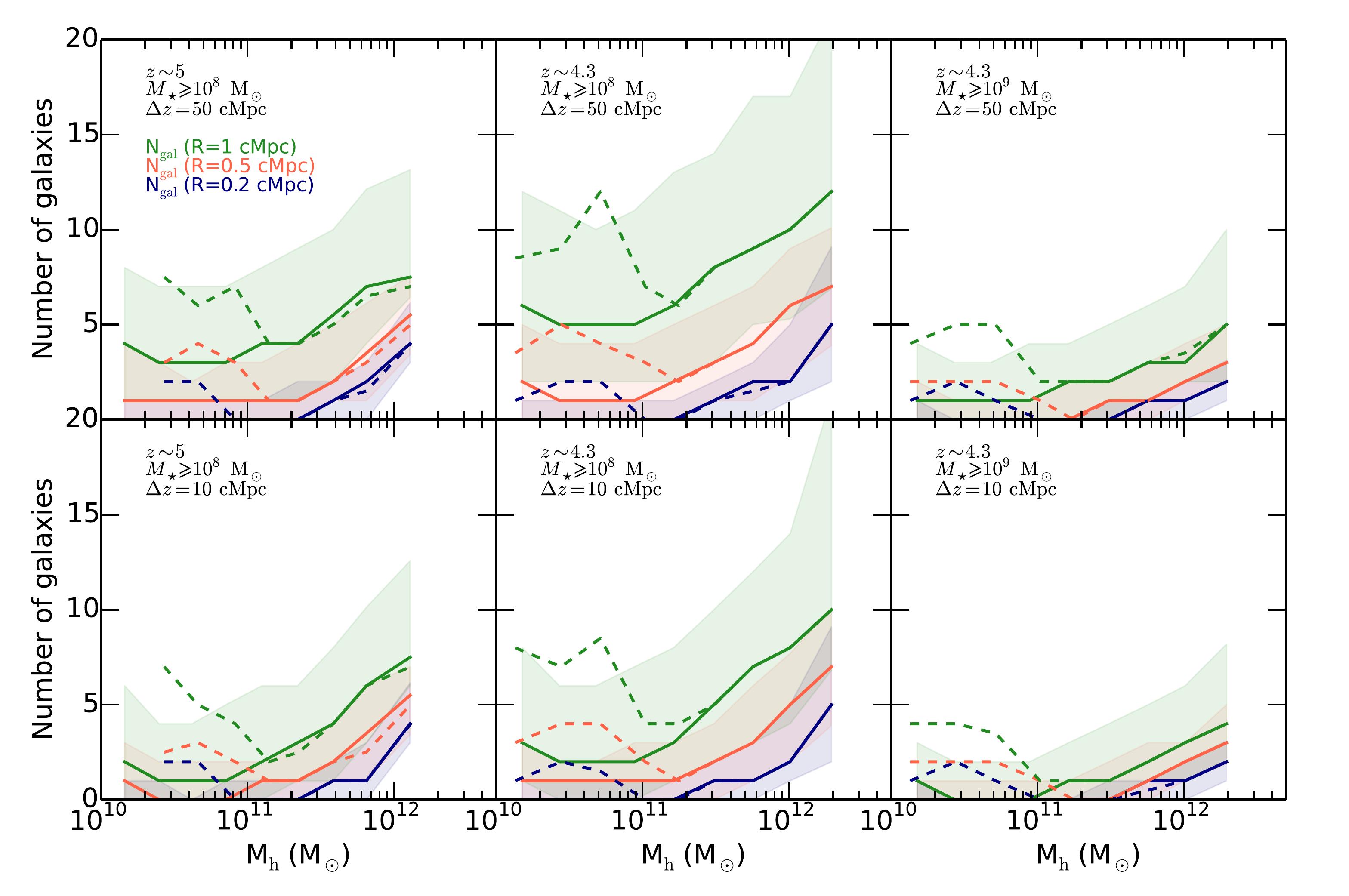}
\caption{Median of the galaxy number counts (at a given projected distance $R$) in the field of view of galaxies as a function of their dark matter halo mass, at $z\sim 5$ (left panels) and $z\sim 4.3$ (middle and right panels). Solid lines represent the median of all the simulated galaxies, and the dashed lines the median for galaxies hosting a BH with $M_{\rm BH}\geqslant 10^{7}\, \rm M_{\odot}$. We either apply a galaxy cut of $M_{\rm \star, thres}=10^{8}\, \rm M_{\odot}$ (left and middle panels), or $M_{\rm \star, thres}=10^{9}\, \rm M_{\odot}$ (left panels). 
Top panels use a projected distance of $\Delta z=50 \, \rm cMpc$, and bottom panels a better redshift accuracy with $\Delta z=10 \, \rm cMpc$.
Shaded areas represent the standard deviation of the distributions $\pm \sigma$. 
For galaxies we find that the number of close galaxies increases for increasing mass haloes. For galaxies hosting a BH, we find the same trend for high mass haloes ($M_{\rm h}>10^{11}\, \rm M_{\odot}$), and a large enhancement of the number of close galaxies for lower mass haloes ($M_{\rm h}<10^{11}\, \rm M_{\odot}$) with respect to the other galaxies.}
\label{fig:halo}
\end{figure*}

\subsection{Star-forming properties of the galaxies in the field of view of massive BH host galaxies}
In this section we investigate the star-forming properties of the galaxies in the field of view of the massive BH host galaxies. We only consider the most massive BHs of $M_{\rm BH}\geqslant 10^{8}\, \rm M_{\odot}$ and use the same selection criteria as in the left panels of Fig.~\ref{fig:nb_gal_radius_most_clustered_lumcut}, i.e. $M_{\star, \rm thres}=10^{8}\, \rm M_{\odot}$ to select galaxies in the field of view and redshift $z\sim5$ and $z\sim4.3$. 
We start by defining the main star forming sequence of the galaxies at these two redshifts, by fitting linearly the SFR of galaxies in the mass range $10^{9}-10^{10}\, \rm M_{\odot}$, which is not much impacted by the presence of quenched galaxies (galaxies with reduced SFR sustained in time). In this section we use SFRs that are averaged over 100 Myr. The main sequence can be defined as:
\begin{eqnarray}
\log_{10} {\rm SFR_{MS}} = \alpha + \beta \times \log_{10} \left( \frac{M_{\star}}{10^{10}\, \rm M_{\odot}} \right) ,
\end{eqnarray}
with $\alpha=1.39,1.22$ and $\beta=0.89,0.85$ for $z\sim5, 4.3$. The relation between SFR and galaxy stellar mass is relatively tight for those high redshifts. Most galaxies are on the main star-forming sequence with a small scatter. Galaxies in the stellar mass range $\log_{10} M_{\star}/\rm M_{\odot}=8-8.5$ have $\rm SFR=0-3 \, M_{\odot}/yr$, galaxies with $\log_{10} M_{\star}/\rm M_{\odot}=8.5-9$ generally have $\rm SFR=0-8 \, M_{\odot}/yr$, galaxies with $\log_{10} M_{\star}/\rm M_{\odot}=9-9.5$ have $\rm SFR=1- 15 \, M_{\odot}/yr$, and more massive galaxies can have SFR of a few tens $\rm M_{\odot}/yr$.

We report in Table~\ref{table_sfr} the median values $\rm \langle SFR \rangle$ and the 15th-85th percentiles of the SFR (in $\rm M_{\odot}/yr$) of the galaxies in the field of view of massive BHs at $z\sim5$ within separation distances of $\rm 0.05\, cMpc$ (second column) and $\rm 1 \, cMpc$ (third column). We show the SFR values for the 5 massive BH environments at $z\sim 5$. Given the large number of BH environments at $z\sim 4.3$, we only provide in the table the five environments with the highest SFR to give an idea of the most star-forming systems.
The median SFR of galaxies in the BH fields has a large variance, and galaxies closer to the BHs often have higher SFR.
The three first rows of the $z\sim 4.3$ section of Table~\ref{table_sfr} show that some nearby galaxies can have very high SFR of $\rm SFR\sim 20\, M_{\odot}/yr$ (corresponding to the most star-forming galaxies for their respective galaxy stellar mass).
In the last column of the table, we quantify how many galaxies (within separation distances of 1 cMpc) in the BH fields of view have SFR above the star-forming sequence of the simulation (i.e., with $\rm SFR\geqslant SFR_{MS}$, as defined in Eq. 3). 
About $\sim 20-60\%$ (at $z\sim 5$) and $0-80\%$\footnote{This large variance can not be seen in Table~\ref{table_sfr}, in which we only includes the five most star-forming environments at $z\sim 4.3$ for illustration of our results.}  
(at $z\sim 4.3$) of galaxies in the field of view of massive BHs are above the star-forming sequence. There is a pretty large variance in the number of star-forming galaxies above the main-sequence. At $z\sim 5$, the median percentage of galaxies above the main-sequence is $29\%$ ($24-44\%$ for the 15th-85th percentiles), and $42\%$ at $z\sim4.3$ ($17-58\%$ for the 15th-85th percentile). 
Our results show that at least a few of the massive BHs environments have more than half of their galaxies forming stars efficiently. \\

\indent In the observations, SFR of $\sim 100\, \rm M_{\odot}/yr$ have been found in companion galaxies of high-redshift quasars \citep{2017Natur.545..457D,2017ApJ...836....8T}. These values are much higher than the tens of $\rm M_{\odot}/yr$ that we find in the simulation. However, the galaxies that are observed have dynamical mass in the range $\log_{10} M/\rm M_{\odot}=11-11.5$ and probe much more extreme quasar environments than the present simulation, for which the stellar mass of the companion galaxies producing high SFR are only in the range $\log_{10} M_{\star}/\rm M_{\odot}=8.3-9.6$.


\begin{table*}
\caption{Star-forming properties of the galaxies in the environment of massive BH host galaxies. We only consider BHs of $M_{\rm BH}\geqslant 10^{8}\, \rm M_{\odot}$ at $z\sim 5$ (5 BHs) and $z\sim 4.3$ (33 BHs) in this table (second column). In the 3rd and 4th column we show the median values and the 15th-85th percentiles of the SFR (in $\rm M_{\odot}/yr$, written as $\rm \langle SFR\rangle_{15th}^{85th}$) of the galaxies in the field of view within separation distances of $\rm 0.05\, cMpc$ and $\rm 1 \, cMpc$. We indicate in the last column the percentage of galaxies within $\rm 1 cMpc$ with SFR above the main star forming sequence of galaxies. For $z\sim 4.3$, we only show the five BH environments with the highest SFR values.}
\begin{center}
\begin{tabular}{ccccc}
\hline
redshift & $M_{\rm BH} \, \rm (M_{\odot})$ & $\rm \langle SFR \rangle$ within 0.05 cMpc& $\rm \langle SFR \rangle$ within 1 cMpc & Percentage above MS (1 cMpc)\\
\hline
\hline
$z\sim 5$ & $1.6\times 10^{8}$ & $3.9_{2.5}^{5.3}$ & $4.5_{1.5}^{5.9}$ & $60\%$ \\
& $1.9\times 10^{8}$ & $8.0_{6.0}^{10.1}$ & $3.4_{0.4}^{11.7}$ & $29\%$ \\
& $1.4\times 10^{8}$ & $1.0_{0.6}^{1.5}$ & $1.7_{0.9}^{2.1}$ & $29\%$ \\
& $2.2\times 10^{8}$ & $0.5_{0.4}^{0.7}$ & $1.4_{0.6}^{2.8}$ & $18\%$ \\
& $2.2\times 10^{8}$ & $0.2_{0.2}^{0.2}$ & $1.6_{0.5}^{13.0}$ & $33\%$ \\
\hline
$z\sim 4.3$ & $2.6\times 10^{8}$ & $14.1_{12.6}^{15.5}$ & $0.86_{0.1}^{9.4}$ & $56\%$ \\
& $3.8\times 10^{8}$ & $10.9_{10.9}^{10.9}$ & $1.8_{0.7}^{3.1}$ & $38\%$ \\
& $1.8\times 10^{8}$ & $10.7_{4.5}^{16.9}$ & $3.4_{0.9}^{8.7}$ & $50\%$ \\
& $1.5\times 10^{8}$ & $5.3_{5.1}^{5.5}$ & $2.3_{0.3}^{5.3}$ & $38\%$ \\
& $1.1\times 10^{8}$ & $3.4_{1.1}^{5.7}$ & $1.0_{0.7}^{3.2}$ & $55\%$ \\
\hline
\end{tabular}
\end{center}
\label{table_sfr}
\end{table*}

\subsection{Number counts of galaxies as a tracer of the halo overdensity}
In this section we investigate the correlation between the number counts of galaxies in a given field of view around a massive BH, and the mass of its host dark matter halo. Fig.~\ref{fig:halo} shows the median of the galaxy number counts in a field centered on all given haloes present in the simulated volume as a function of the virial mass of these haloes as solid colored lines. The galaxy number counts in the field of massive BHs ($M_{\rm BH}\geqslant 10^{7}\, M_{\odot}$) are shown as dashed lines. 
Colors indicate different 2D projected radii within which we compute the galaxy number counts, i.e. $R=0.2, 0.5, 1\, \rm cMpc$. We show the same figure with higher apertures in Fig.~\ref{fig:halo_higher_apertures}, and find similar results.
The left panels are for redshift $z\sim5$, and the middle and right panels for $z\sim4.3$. 
The standard deviations $\pm \sigma$ are shown in shaded areas. For the dashed lines, bins below $M_{\rm h}=10^{11}\, \rm M_{\odot}$ (for both $z\sim 5$, and $z\sim 4.3$) and higher than $M_{\rm h}\sim10^{11.7}\, \rm M_{\odot}$  for $z\sim 5$ have the lowest statistics with less than 50 BHs, and therefore larger uncertainties. 
In the left and middle panels, we apply a selection in galaxy stellar mass $M_{\star}\geqslant 10^{8}\, \rm M_{\odot}$. In the right panels, we test how the correlation between halo mass and number counts evolves with a more restrictive selection of galaxies, by selecting only galaxies with $M_{\star}\geqslant 10^{9}\, \rm M_{\odot}$ when computing the galaxy number counts.
In the top panels we compute the number counts of galaxies assuming a projected distance of $\Delta z = 50\, \rm cMpc$. In the bottom panels, we assume $\Delta z = 10\, \rm cMpc$.

The first thing to notice is that, as expected, the galaxy number counts within a given radius around a galaxy trace the mass of the dark matter halo of this galaxy. In other words, more galaxies are located in the field of more massive haloes. In Fig.~\ref{fig:haloes_HnoAGN} we compare the Horizon-AGN and Horizon-noAGN simulations and we find very similar results. The median of the galaxy number counts is slightly enhanced for the most massive haloes in the simulation without AGN feedback.

Now, when we look at the number of galaxies in the field of view of massive BHs, i.e. the dashed lines, we find almost no difference with the global behavior for massive haloes of $M_{\rm h}\gtrsim 2\times 10^{11}\, \rm M_{\odot}$.
However, we find a different behavior for lower mass haloes, where the number counts of galaxies around massive BHs is much higher on average. 
At $z\sim 5$, we find a median of the BH-halo mass ratio $M_{\rm BH}/M_{\rm h}=8.25\times 10^{-5}$, and a mean of $M_{\rm BH}/M_{\rm h}=9.5\times 10^{-5}$ for BHs with $M_{\rm BH}\geqslant 10^{7}\, \rm M_{\odot}$. For the very few cases of more massive BHs of $M_{\rm BH}\geqslant 10^{8}\, \rm M_{\odot}$, we find slightly higher values with a median and mean of $M_{\rm BH}/M_{\rm h}=1.2\times 10^{-4}\, \rm M_{\odot}$. Very similar values are found at $z\sim 4.3$. We find that there are some BHs with mass significantly larger than expected from the mean ratio $M_{\rm BH}/M_{\rm h}$, with a maximum mass ratio of $M_{\rm BH}/M_{\rm h}=0.004$ for BHs with $M_{\rm BH}\geqslant 10^{8}\, \rm M_{\odot}$.

The right panel is identical to the middle one ($z\sim4.3$), except that we now only select galaxies with $M_{\star}\geqslant 10^{9}\, \rm M_{\odot}$. As expected the median of the galaxy number counts is lower, and the correlation between the number count and dark matter halo mass lower as well. However we still find an enhancement of the galaxy number counts in the field of view of massive BHs of $M_{\rm BH}\geqslant 10^{7}\, \rm M_{\odot}$ for low mass haloes of $M_{\rm h}\leqslant 10^{7}\, \rm M_{\odot}$.


To conclude, we confirm that the galaxy number counts around a given galaxy statistically correlate with the dark matter halo mass of its host halo, as expected. Interestingly, we find that the number counts of galaxies around massive BHs of $M_{\rm BH}\geqslant 10^{7}\, \rm M_{\odot}$ in haloes of $M_{\rm h}\leqslant 10^{11}\, \rm M_{\odot}$ is enhanced compared to other haloes. The even more massive BHs of $M_{\rm BH}\geqslant 10^{8}\, \rm M_{\odot}$ present in the simulation are located in more massive haloes of $M_{\rm h}>10^{11}\, \rm M_{\odot}$; for which we do not find a galaxy count enhancement on average compare to haloes without BHs. 
Finally, we find very similar results when the redshift estimate is better, i.e. with $\Delta z = 10\, \rm cMpc$.
The galaxy counts around massive BHs does not appear to strongly constrain the halo mass, independently of the redshift precision of the galaxies.

\subsection{Uncertainties related to the sub-grid physics models}
At the resolution of the simulation Horizon-AGN (and the other large scale simulations, or even zoom-in galaxy simulations) subgrid models are needed  to model physical processes taking place at the galactic scale or below.

Our present study is based on the number counts of galaxies in the field of view of the massive BHs in our samples. The subgrid physics for BH and galaxy evolution can affect both the conception of our samples (the massive BHs could be more or less massive) and the number counts of galaxies.
The subgrid galaxy models can also affect the number counts of galaxies: the impact of the supernova feedback model has been discussed in \citet{Costa:2013aia}, and we discuss in details the impact of AGN feedback in the next section \citep[see also][]{Costa:2013aia}.
\citet{Costa:2013aia} show that SN feedback can impact the number counts of galaxies in the quasar field of view, by reducing the star formation rate of the galaxies for stronger winds, and so the number of detectable galaxies.
BH seeding, accretion or feedback could affect the mass of the BHs that we include in our sample. We discuss the uncertainties of our models in the following.

A simple model for BH seeding is used in the simulation. The occupation fraction of galaxies, i.e. the number of galaxies hosting a BH, is predicted to vary for different models of BH formation \citep[light vs heavy seed models,][]{2008MNRAS.383.1079V,2012NatCo...3E1304G}. The number of newly formed BHs through different formation models and their BH-BH merger history (depending on the efficiency of the BH formation model) could impact the mass of the BHs at $z=6$. Unfortunately using several models of BH formation to run such large cosmological simulations is not feasible yet.
Changing the initial mass of the BHs in our simulation should not strongly affect our results as we are looking at the most massive BHs \citep[see discussions in ][]{Sijacki2009,Costa:2013aia}, i.e. BHs that have experienced rapid growth. Their masses at $z=5-6$ are already more than one order of magnitude higher than the seeding mass ($M_{\rm BH}=10^{5}\, \rm M_{\odot}$, or less if not enough available gas). 

The modeling of accretion onto the BHs can also affect their growth history and their mass at $z\sim 6$. Here, we have used the Bondi-Hoyle-Lyttleton accretion rate model, capped at the Eddington accretion rate. Different models, such as the gravitational torque model, have been employed in galaxy merger simulations \citep{2010MNRAS.407.1529H} or large-scale cosmological simulations \citep{2017MNRAS.464.2840A}. The accretion model used in simulations affects the slope and the normalization of the scaling relation \citep{2017MNRAS.464.2840A}. In zoom-in simulations coupled to analytic BH growth models, \citet{2013ApJ...770....5A} show that the two Bondi-Hoyle-Lyttleton and torque models provide different accretion rate history. They find that the Bondi accretion rates can be lower at early times and higher at later times than the accretion rates derived with the torque model. 
Since we studied massive BHs in our analysis, their masses should not be strongly affected by the accretion model.

Finally, it is possible that the low resolution of cosmological simulations, which is not sufficient to accurately resolve gas flows in the close surroundings of BHs, overestimate the accretion rate on small scale \citep{2010MNRAS.407.1529H}. New simulation method allowing to reach a higher resolution in the close vicinity of BHs in cosmological simulations will give us more answers \citep[][Angl\'es-Alc\'azar, in prep]{2019MNRAS.483.3488B}.
Episodes of super-Eddington accretion are predicted to boost the rapid growth of BHs \citep{Volonteri2005}. Allowing for these episodes in the simulation could also alter the mass of the BHs at $z\sim 6$ in our simulation.

\section{Testing the impact of AGN feedback}
It has been suggested that feedback from the quasar can suppress star formation in neighbouring galaxies \citep{1992MNRAS.256P..43E,1996ApJ...465..608T,2002MNRAS.333..177B,2007ApJ...663..765K,2010ApJ...721.1680U,2018arXiv180208912O} and therefore a region a priori overdense is observed to be devoid of star-forming galaxies. This is particularly relevant for studies that select galaxies in UV or via UV-driven lines, e.g., LAEs. Also LAEs tend to be lower-mass galaxies, and could be more affected by radiation from the quasars than LBGs \citep[e.g.,][]{2007ApJ...663..765K,2018arXiv180208912O}.

\subsection{Horizon-AGN and Horizon-noAGN}
The twin large-scale simulations Horizon-AGN and Horizon-noAGN offer us an opportunity to investigate statistically the specific effect of AGN feedback on the number counts of galaxies around quasars. The galaxy mass function of the two simulations at $z=5$ is in good agreement with observations \citep[][Fig. 7]{2017MNRAS.467.4739K}, but at later time Horizon-noAGN shows an increasing overestimate, similar to what is found in the stellar-halo mass relation \citep[][Fig. 10]{2016MNRAS.463.3948D}. Including AGN feedback leads to less massive/luminous galaxies, so that at a given luminosity, there are fewer galaxies in the simulation Horizon-AGN. We can then expect the galaxy counts around massive BHs to be lower, at least at large scale, resulting from the {\it global effect} that there are less luminous galaxies in the simulation box (because of their own AGN feedback). The counts at very small scales are more likely to be representative of the impact of a given BH on its surroundings, and we discuss this in more detail in the Section 5.2.

We carry out object-to-object comparisons to ensure that we study the impact of AGN feedback on the same objects in the two simulations. We follow the method developed in \citet{2017MNRAS.472.2153P}. Each galaxy is first coupled to a dark matter halo. Galaxy-halo pairs are chosen by associating the most massive galaxy located within $5\%$ of the virial radius of its host halo. Dark matter haloes are matched initially between the two simulations if $75\%$ or more of the dark matter particles in one given halo of Horizon-AGN are also present in the corresponding halo in Horizon-noAGN. To exclude spurious couples of dark matter haloes, we only consider couples with total halo virial mass ratio higher than $0.1$ and lower than $10$. 
For example at $z\sim5$ we match $61\%$ of Horizon-AGN galaxies hosting BHs of $M_{\rm BH}>10^{7}\, \rm M_{\odot}$ with galaxies in Horizon-noAGN , $84\%$ for BHs of $M_{\rm BH}>6\times10^{7}\, \rm M_{\odot}$, and $80\%$ for BHs of $M_{\rm BH}>10^{8}\, \rm M_{\odot}$. Here we need to remember that even if we can connect galaxies from one simulation to another, the processes taking place in those galaxies are still different, on potentially different timescales, and/or shifted in time. However, as we here focus our comparison to the counts around massive BHs, i.e. on the number counts of galaxies of a given luminosity, we do not expect this potential shift in time of the evolution of galaxies between the two simulations to affect much our comparison.

%

\begin{figure*}
\centering
\includegraphics[width=\textwidth]{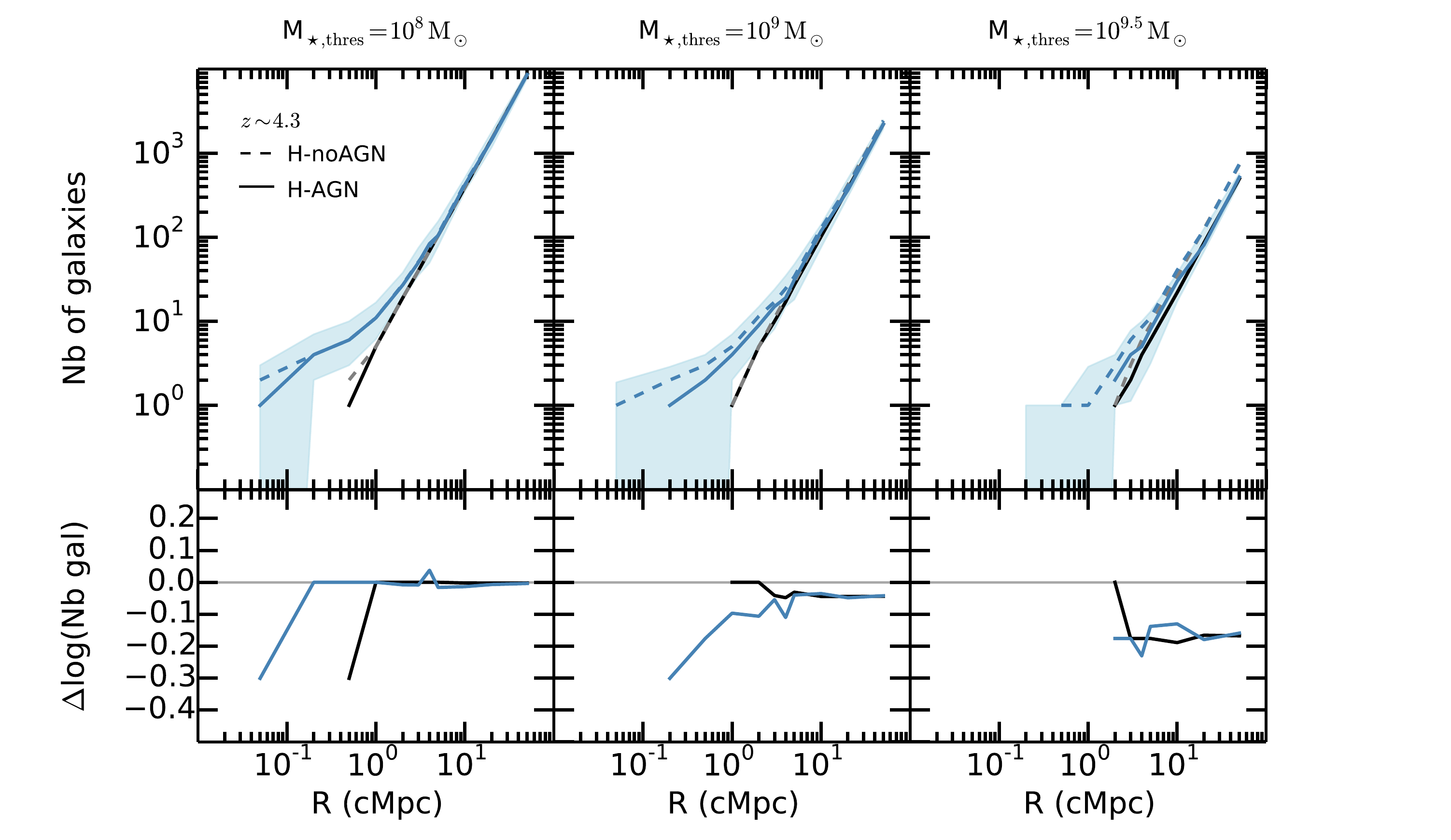}
\caption{Galaxy number counts around BHs in Horizon-AGN at $z\sim 4.3$ (solid lines), and around the corresponding galaxies in Horizon-noAGN (dashed lines), for 3 different stellar mass thresholds $M_{\rm \star, thres}=10^{8}\,\rm M_{\odot}$ (left panels), $M_{\rm \star, thres}=10^{9}\, \rm M_{\odot}$ (middle panels),  and $M_{\rm \star, thres}=10^{9.5}\, \rm M_{\odot}$ (right panels). For clarity, we only show BHs with $M_{\rm BH}>10^8 \, M_{\odot}$, as blue solid and dashed lines, but the two other ranges of BH mass have similar behavior. The blue shaded region represents the standard deviation $\pm 1\sigma$ for the simulation Horizon-AGN. 
Black and dashed grey lines represent the number counts of galaxies averaged over all the simulated galaxies for Horizon-AGN and Horizon-noAGN, respectively, and are shown here for reference.
Bottom panels show the residual of the logarithm of the number of galaxies of the simulation Horizon-AGN relative to that of the simulation Horizon-noAGN.}
\label{fig:HAGNnoAGN_nb_gal_radius_lumcut}
\end{figure*}

In Fig.~\ref{fig:HAGNnoAGN_nb_gal_radius_lumcut}, we show the average galaxy number counts of the entire population of galaxies in black solid lines for the simulation Horizon-AGN, and in grey dashed lines for the simulation Horizon-noAGN, at $z\sim 4.3$. As before we use three different stellar mass thresholds to count galaxies:  $M_{\rm \star, thres}=10^{8}\, \rm{M_{\odot}}$ (left panel), $M_{\rm \star, thres}=10^{9}\, \rm{M_{\odot}}$ (middle panel), and $M_{\rm \star, thres}=10^{9.5}\, \rm{M_{\odot}}$ (right panel). The average galaxy number counts (averaged over all galaxies in the simulated volume) are very similar in Horizon-AGN and Horizon-noAGN.
We show the galaxy number counts in the field of view of $M_{\rm BH}\geqslant 10^{8}\, \rm M_{\odot}$ BHs as blue solid (Horizon-AGN) and dashed lines (matched galaxies in Horizon-noAGN). To keep the figure clear, we only show this BH mass range, but we obtain similar results (with lower amplitude) for ranges of lower BH mass.
The median of the galaxy number counts in BH fields is very similar in the simulations Horizon-AGN and Horizon-noAGN. The number counts are only very slightly higher in Horizon-noAGN when considering a low stellar mass limit of $M_{\rm \star, thres}=10^{8}\, \rm M_{\odot}$. The deviation slightly increases for more restrictive selections of galaxies, while still being statistically not significant.
We find that AGN feedback, either from the massive BH host galaxies or from the galaxies in the field of view, is statistically only slightly impacting the galaxy number counts when fainter galaxies can be observed. However, for more restrictive selections of galaxies (i.e. only more massive galaxies), AGN activity in the neighboring galaxies can make their observational detections harder because of suppressed star formation, masking the potential overdensity of the regions.

To illustrate our findings, we present in Fig.~\ref{fig:density_maps} two projected gas density maps of the simulations Horizon-AGN (top panel), and Horizon-noAGN (bottom panel), centered on the same galaxy, indicated with a pink cross symbol. 
White circles indicate neighborhing galaxies with stellar mass of $M_{\star}\geqslant 10^{8} \, \rm M_{\odot}$, and grey circles galaxies with $M_{\star}\geqslant 10^{9.5} \, \rm M_{\odot}$. For Horizon-AGN, galaxies hosting a central BH are shown with a white or grey cross. We see here by comparing the two gas density maps, that fewer bright/massive galaxies (grey circles) are indeed found in the top panel for the simulation Horizon-AGN than for the simulation Horizon-noAGN.

\begin{figure}
\centering
\includegraphics[width=\columnwidth]{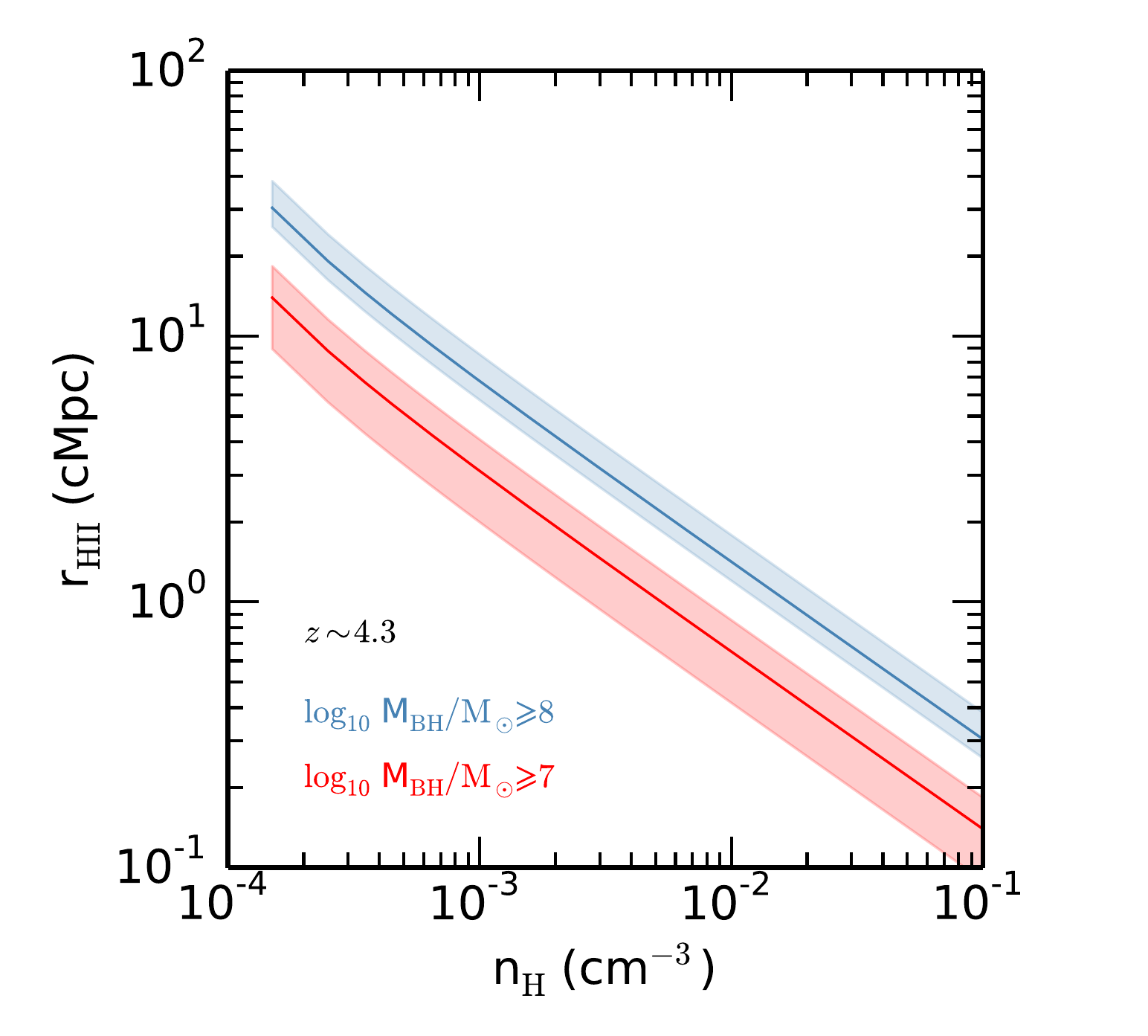}
\caption{Median size of HII region around massive BHs of $M_{\rm BH}\geqslant 10^{8} \, \rm M_{\odot}$ (blue solid line), and of  $M_{\rm BH}\geqslant 10^{7} \, \rm M_{\odot}$ (red solid line) in Horizon-AGN at $z\sim 4.3$. Shaded areas represent the $\pm 1\sigma$ of the distributions.}
\label{fig:HII_region}
\end{figure}

\subsection{Radiation from massive BHs?}
In the previous section we have seen that the impact of AGN feedback was very weak, and mostly due to the presence of AGN in the nearby galaxies rather than a real effect from the massive BH host galaxies we were interested in. In the present section we want to estimate on what scales radiation feedback, which is not included in Horizon-AGN, might affect galaxies.
When studying the quasar CFHQS J2329-0301, \citet{2010ApJ...721.1680U}  find that the galaxies in the field of view of the quasar are concentrated in a ring-like shape of radius $\sim 3\rm pMpc$ (i.e. $\sim 15\rm cMpc$). Similarly, \citet{2007ApJ...663..765K} find that there is no LAEs galaxies within $4.5\, \rm cMpc$ around the quasar QSO SDSS J0211-0009 at $z=4.87$ (though several LBG are identified, but with larger uncertainties in the determination of their redshift).
AGN feedback from the central quasar has been invoked to explain the absence of galaxies within this ring-like region. We have seen in the previous section that the thermal component of AGN feedback is not sufficient to greatly affect the properties of galaxies around a quasar. The possibility remains that strong radiation emanating from the quasar, which is not included in Horizon-AGN, could photoionize the gas in the galaxies present in its close vicinity and quench their star formation.
As described in section 4.2, close galaxies and satellite galaxies, i.e. galaxies residing in the same dark matter haloes, which are located in the close vicinity of massive BHs, can be impacted by strong radiation from these central massive BHs.
Radiation-hydrodynamical simulation are necessary to investigate the effect of strong radiation on the neighboring environment of quasars, and demonstrate whether they are able to cease star formation in the surrounding galaxies, and on which length scale. 
This is beyond the scope of the present paper. Instead, we follow another approach, and compute the HII region radius around all the simulated BHs in Horizon-AGN.  

The size of a HII region around a BH, depends on the ambient density of hydrogen ($n_{\rm H}$, in $\rm cm^{-3}$), and the rate of ionizing photons ($Q_{\star}$, in $\rm s^{-1}$) that are emitted by the BH. The radius of the HII region can be expressed as:
\begin{eqnarray}
r_{\rm HII}=\left(\frac{Q_{\star}}{R_{\rm rec}}\frac{3}{4\pi}\right)^{1/3}=\left(\frac{Q_{\star}}{\alpha_{\rm T}\, n_{\rm H}^{2}}\frac{3}{4\pi}\right)^{1/3} ,
\end{eqnarray}
where the recombination rate per unit volume is defined by $R_{\rm rec}=\alpha_{\rm T} \, n_{\rm H}\, n_{\rm e}=\alpha_{\rm T} \, n_{\rm H}^{2}$, and $\alpha_{T}=2.6 \times 10^{-13} \rm cm^{3}s^{-1}$ is the temperature dependent radiative recombination coefficient. Because hydrogen is the most abundant element and is fully ionized in the region, the volume density of hydrogen and electron can be considered equal ($n_{\rm H}= n_{\rm e}$).

In Fig.~\ref{fig:HII_region}, we show the size of the HII regions around BHs as a function of the volume density of hydrogen $n_{\rm H}$. This figure illustrates the dependency of $r_{\rm HII}$ with $n_{\rm H}$, where we compute the ionizing photon rate $Q_{\star}$ from each BH of the simulation, i.e. self-consistently taking BH mass and BH Eddington ratio from the simulation.   
We show the median size of the HII region around massive BHs of $M_{\rm BH}\geqslant 10^{7} \, \rm M_{\odot}$ in red solid line, and of  $M_{\rm BH}\geqslant 10^{8} \, \rm M_{\odot}$ in blue solid line, at $z\sim 4.3$. Shaded areas represent the standard deviation $\pm 1\sigma$ of the distributions.

\begin{figure}
\centering
\includegraphics[scale=0.48]{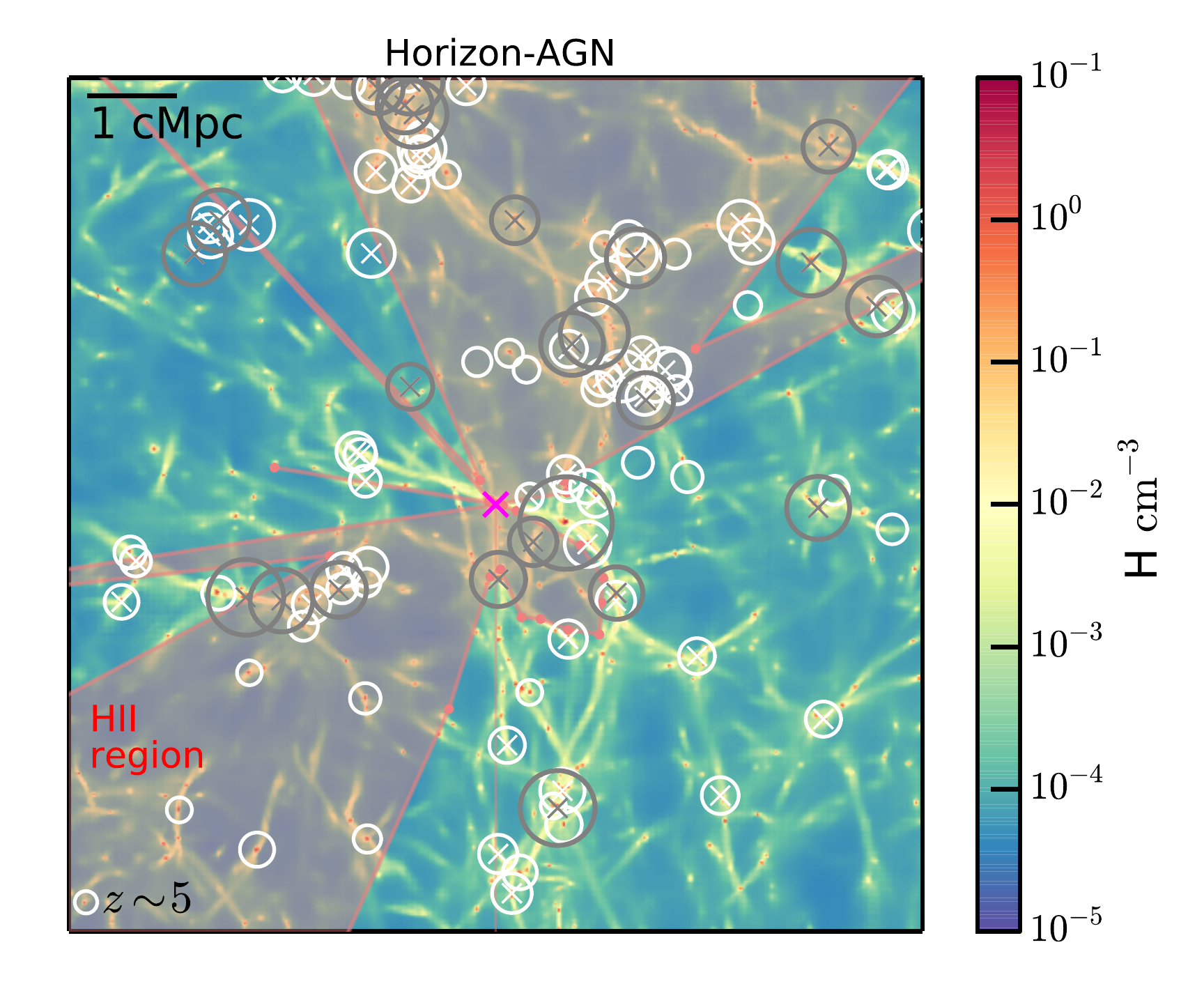}
\includegraphics[scale=0.48]{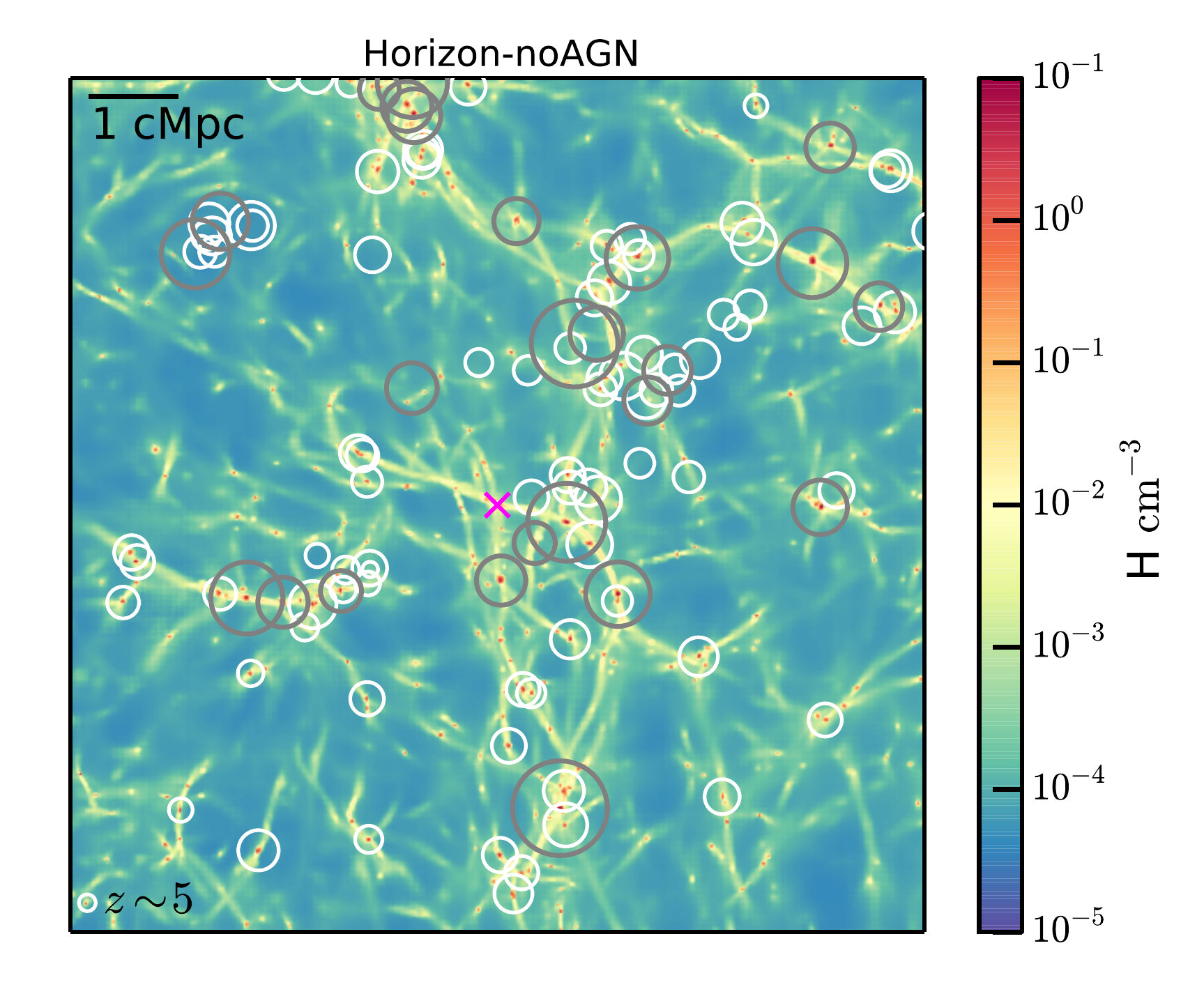}
\caption{Projected gas density maps (in unit of $\rm H/cm^{3}$) of the simulations Horizon-AGN (top panel), and Horizon-noAGN (bottom panel), centered on the same galaxy (indicated by a pink cross) hosting a BH with $M_{\rm BH}=6\times 10^{7}\, \rm M_{\odot}$. White circles indicate neighborhood galaxies with $M_{\star}\geqslant 10^{8} \, \rm M_{\odot}$, and grey circles galaxies with $M_{\star}\geqslant 10^{9.5} \, \rm M_{\odot}$. The circle radii are equal to 10 times the virial radius of galaxies. For Horizon-AGN, galaxies hosting a central BH are shown with a cross. AGN feedback lowers star formation rate and the resulting luminosity/stellar mass in massive galaxies hosting a BH. Indeed we see here by comparing the two gas density maps, that fewer massive galaxies (grey circles) are found in the top panel for the simulation Horizon-AGN than for the simulation Horizon-noAGN.
On the top panel, we show the HII region of the BH in the plane (x,y). Pink dots indicate $r_{\rm HII}$ computed along 40 lines of sight in this plane.}
\label{fig:density_maps}
\end{figure}

To illustrate the shape of HII regions, we also provide on the top panel of Fig.~\ref{fig:density_maps} the size of the HII region (pink shaded region) for an individual BH in the simulated volume. This BH has a mass of $M_{\rm BH}=6\times 10^{7}\, \rm {M_{\odot}}$ and an Eddington ratio of $\log_{10} f_{\rm Edd}=-1$, at $z\sim 5$.
We first compute the $n_{\rm H}$ profile along 40 different lines-of-sight, all in the plane (x,y) (i.e. the plane represented in Fig.~\ref{fig:density_maps}), and in radial bins from 0 to 5 cMpc. The minimum and maximum $n_{\rm H}$ values of these 40 profiles are presented in Fig.~\ref{fig:profile} in a blue shaded area. For comparison, we also show the mean $n_{\rm H}$ profile as a grey line, computed in spherical shells (i.e. in 3D space) from 0 to 5 cMpc around the BH. 
We then numerically solve Eq. 5 for a non-homogenous $n_{\rm H}$ medium using the $n_{\rm H}$ profiles along the lines-of-sight. Resulting $r_{\rm HII}$ values for all lines-of-sight are shown as pink dots in Fig.~\ref{fig:density_maps}, highlighting the high anisotropy of the HII region around the BH. The HII region extends to large radii (with $r_{\rm HII} > 5\, \rm cMpc$) in some directions. In some others the HII region size is much more reduced due to very dense environments. This is the case for the bottom left region around the BH where we note the presence of many galaxies in the plane (x,y).

Impacted regions can have a size from below $1\, \rm cMpc$ to $\sim 10\, \rm cMpc$ \citep[see also][]{2005ApJ...628..575W}, depending on the BH mass, its accretion rate, and the hydrogen density of its environment. These are the scales that are probing the enhancement of the number counts, see Section 4. The ability of this strong radiation to cease or alter star formation in these nearby galaxies, and consequently their detectability, also depends on their own properties, e.g. their mass. Low-mass galaxies would be more impacted \citep{1986MNRAS.218P..25R,1992MNRAS.255..346B,1992MNRAS.256P..43E}.
\citet{2007ApJ...663..765K} qualitatively estimated that UV radiation from quasars can suppress galaxy star formation in haloes with $M_{\rm vir}<10^{9}\, \rm M_{\odot}$, but would not affect more massive haloes of $M_{\rm vir}>10^{11}\, \rm M_{\odot}$. 
\citet{Costa:2013aia} investigate this effect by simulating several overdense quasar environments with increasing strength of the UV background.  The enhanced UV background is set to mimic stronger photo-heating from the quasar(s) and the more numerous star-forming galaxies in the regions, but is not computed self-consistently as a function of these objects in the simulations. As discussed below, the photo-ionizing radiation from a quasar could be highly non-isotropic in space, and impact only galaxies/halos in preferred directions depending on the gas and large-scale distribution.
They find that even with a stronger ionizing flux by a factor 100 the star formation of the surrounding halos with $M_{\rm h}\geqslant 10^{9}\, \rm M_{\odot}$ is barely impacted. Star formation is only affected in low-mass halos of $M_{\rm h}<10^{9}\, \rm M_{\odot}$ \citep{Costa:2013aia}, as suggested by \citet{2007ApJ...663..765K}.

This can have strong consequences on the distribution and type of galaxies around quasars. LBGs with mass $M_{\rm vir}\sim 10^{11-12}\, \rm M_{\odot}$ would not be affected by radiation, but LAEs, which are believed to be less massive in general, would be more affected. Another feedback of massive BHs could be radiation pressure-driven winds. These winds could impact even more the nearby galaxy satellites by sweeping out or evaporating their gas content.
The duty cycle of quasars is also an important parameter, strong activity of the quasars at early times would delay star formation in the surroundings, preventing their observability. However, if the activity of the BHs peaks after the formation of the surrounding galaxies, the galaxy enhancement could still be observable. 
The duty cycle of quasars is constrained by comparing the number density of quasars with the number density of dark matter haloes \citep{2001ApJ...547...27H,Martini2001,2008MNRAS.390.1179W,2011ApJ...735..117F,2017ApJ...840...24E}, but constraints remain limited at high redshift.
Several other processes could explain the ring-like shape of the galaxy density around quasars, as well as observation biases if low-mass galaxies are missed (see end of Section 4.2).
Therefore further theoretical studies using cosmological simulations of massive galaxies with massive BHs using radiative transfer to study the impact of BH feedback on surrounding small galaxies (with e.g. self-consistent filling factor of the medium, escape fraction, etc) are needed to improve our understanding of this observational feature.

\section{Conclusions}
In this paper, we have investigated the galaxy number counts around the most massive BHs in the state-of-the-art large scale cosmological hydrodynamical simulation Horizon-AGN, and compared them to the average number counts computed by averaging over all galaxies in the simulated volume. Our goal was to a) provide a theoretical framework with hydrodynamical simulations to investigate/understand the puzzling diversity of quasar environments that have been found so far in observations (Fig.~\ref{fig:obs_data}) and the environment of fainter BHs that start being observed, and b) to predict how the upcoming space missions JWST, WFIRST, and Euclid will help us improve our understanding of high-z massive BH environments. 
In the following we summarize our main results.

\begin{itemize}
\item We find that there is an excess of 3D galaxy counts in the neighbourhood of massive BHs compared to average field (Fig.~\ref{fig:nb_gal_radius_spherical_projected}). This excess can be up to 10 galaxies within a sphere of radius 1 cMpc at z=5-4, for exemple. This excess is statistically significant for BHs with $M_{\rm BH}\geqslant 10^{8}\, \rm M_{\odot}$ for spherical separation radius of $R\lesssim 10\, \rm cMpc$. The enhancement of the number counts is more pronounced for more massive BHs. 

\item When using 2D projected separation distances, we find that the enhancement of galaxy number counts in the field of view of massive BHs partly vanishes/becomes statistically insignificant (Fig.~\ref{fig:nb_gal_radius_spherical_projected}), and is significant only at very small radius for the most massive BHs ($< 1\, \rm cMpc$) for which we find a difference of a few galaxies. We still find that the enhancement of the number counts is more pronounced for more massive BHs. 

\item In order to mimic the ability of the upcoming missions JWST, WFIRST and Euclid to detect galaxies, we have applied three different stellar mass thresholds to select galaxies present in the BH fields of view (Fig.~\ref{fig:pict_1}). While the high sensitivity of JWST should allow to detect the enhancement of the galaxy number counts around BHs of $M_{\rm BH}\geqslant 10^{8}\, \rm M_{\odot}$, selecting only massive galaxies (with e.g., WFIRST or Euclid sensitivity) decreases the amplitude of the overdensity compared to average field. 

\item Moreover, we also find that the enhancement of the galaxy number counts around the most massive BHs is larger (especially for non-restrictive cuts to select galaxies) for better redshift precision (Fig.~\ref{fig:smaller_length}) of the galaxies in the fields than LAE galaxies, for example with grism measurements. For example, the number counts of galaxies at $z\sim 5$ with a stellar mass cut of $M_{\rm \star,thres}\geqslant 10^{9}\, \rm M_{\odot}$ is significantly enhanced compared to average until 2 cMpc, instead of 1 cMpc with the redshift precision of LAEs.

\item The variance of the number counts around massive BHs is large, and decreases for the environments of the most massive BHs with $M_{\rm BH}\geqslant 10^{8}\, \rm M_{\odot}$ (Fig.~\ref{fig:nb_gal_radius_most_clustered_lumcut}). Using a higher stellar mass cut to select galaxies in the fields of view also increases the variance, even for the most massive BHs. 

\item In the field of view of massive BHs, we find an excess of close companions with small projected separations of $R\leqslant 1\, \rm cMpc$ ($\sim 15-20\, \rm pkpc$ in proper distances at $z\sim 5$). This is in good agreement with recent observational findings \citep{2017ApJ...836....8T,2017Natur.545..457D}.
Interestingly, we find a few cases for which the BH host galaxies have a close galaxy companion with no further overdensity at large scales.

\item We find that at least some of the closest companion galaxies (within separation distances of 0.05 cMpc) of the most massive BH ($M_{\rm BH}\geqslant 10^{8}\, \rm M_{\odot}$) host galaxies in the simulation have high SFR up to $\rm SFR \sim 20\,  \rm M_{\odot}/yr$. Recent observations have found SFR of $\sim 100 \, \rm M_{\odot}/yr$ in companion galaxies \citep{2017Natur.545..457D,2017ApJ...836....8T} but for galaxies with mass of one order of magnitude higher. 


\item As expected, we confirm that there is a correlation between halo masses and galaxy number counts when averaged in the entire simulated volume. 
We find a similar correlation between the mass of massive BH ($M_{\rm BH}\geqslant 10^{7}\, \rm M_{\odot}$) host haloes and galaxy number counts for haloes of $M_{\rm h}\geqslant 10^{11}\, \rm M_{\odot}$. However, we find an enhancement of the galaxy number counts around massive BHs for lower mass haloes. Galaxy counts around massive BHs do not appear to strongly constrain halo mass, even when we increase the redshift precision of the counted galaxies.


\item We find that AGN feedback makes the detection of bright/massive galaxies around an AGN less likely because it suppresses star formation in the most massive galaxies. The effect, based on the implementation in Horizon-AGN, is small (Fig.~\ref{fig:HAGNnoAGN_nb_gal_radius_lumcut}). 
\item Radiative feedback from the quasar, namely the ability to photoionize gas in low-mass galaxies and suppress their ability to form stars, is potentially important. We calculated the theoretical size of the HII regions around massive BHs in Horizon-AGN along several lines-of-sight (Fig.~\ref{fig:density_maps}), and find that they can extend on large scales, $1-10\, \rm cMpc$. Our results also highlight the anisotropy of the HII regions, depending on the environment of the high-z BHs, i.e. on the density of hydrogen along different lines-of-sight emanating from the BHs.  

\end{itemize}

In this paper we find that even if statistically massive BHs live in overdense regions of the Universe, a large diversity of environments is to be expected at high redshift. This large diversity found in the simulation is enhanced by the difficulty of observing high-redshift galaxies, and the uncertainties on galaxy's redshift estimate. 
Therefore the large diversity of environments found in the observations, i.e. so far the environments of about 20 quasars in the range $z\sim4-7$, does not seem unexpected. 
In our analysis we are limited by the volume of the simulation, which does not produce BHs as massive and luminous as the most massive quasars observed at $z\sim 5-6$. Therefore our findings need to be somehow extrapolated to higher mass BHs. However, with BHs of $M_{\rm BH}\sim 10^{7}-10^{8}\, \rm M_{\odot}$ and luminosities of $L_{\rm bol}\sim 10^{46}\, \rm erg/s$ (at $z\sim 5-6$), we probe the regime of the BH population that start to be observed at high redshift \citep{2019arXiv190407278O}.

The environments of massive BHs and the number counts of close companions in the high-redshift Universe provide us with clues on the emergence of these very massive objects (e.g., their growth, role of galaxy mergers), and their possible interplay with surrounding galaxies (e.g., AGN feedback). 
New upcoming space missions such as JWST or WFIRST will have the power to constrain even more the environments of massive BHs, including the regime of BH mass that we cover here. 
The high capacities of JWST to observe fainter galaxies will allow to observe again the environment of some high-redshift quasars, for which an overdensity was not found. For example, there were no galaxy observed in the close vicinity ($\leqslant 3\, \rm pMpc$) of the quasar CFHQS J2329-0301 \citep[$z=6.4$,][]{2010ApJ...721.1680U}. 
AGN feedback from the quasar could have delayed the formation of close galaxies, or suppressed their star formation rate. Looking for fainter galaxies with JWST will help us constrain the role of AGN feedback for these extremely massive objects. 

In addition to detecting lower mass galaxies, we find that the difference between the number counts of galaxies in the field of view of massive BHs and the average number counts is higher for better precision of the redshift estimates. Therefore, we find that detections of LAE galaxies, or even grism measurements, should be more efficient to constrain the possible overdensity of massive BH environments.

One possible explanation for the diversity of quasar environments has been that these quasars would actually not be located in the most massive halos, but in fewer massive halos \citep{Fanidakis:2013uva} for which we should expect less number counts of galaxies.
We have investigated this in our simulation.
In Horizon-AGN, BHs of  $M_{\rm BH}\geqslant10^8 \, \rm M_{\odot}$ at $z=6-4$ are hosted by halos with masses in the range  $M_{\rm h}\sim 3\times10^{11}- 5\times10^{12} \, \rm M_{\odot}$. The upper value corresponds to the mass of the most massive halos of the simulation at that time. The most massive BHs are definitely in the most massive halos in our simulation, i.e. $M_{\rm h}\sim 10^{12}\, \rm M_{\odot}$. This halo mass range is in good agreement with \citep{Fanidakis:2013uva} (based on the GALFORM semi-analytical model). They define quasars as BHs with $L_{\rm bol}>46 \, \rm erg/s$\footnote{BHs of $M_{\rm BH}\geqslant10^8 \, \rm M_{\odot}$ have bolometric luminosities of  $L_{\rm bol}\sim 10^{42-47} \, \rm erg/s$ at $z=6-4$ in Horizon-AGN.}, and find that these quasars do not populate more massive halos. We can not confirm their findings because the volume of Horizon-AGN is not sufficient to have more massive halos than a few $10^{12}\, \rm M_{\odot}$.
Answering this question of the dark matter halo properties of the quasars of $M_{\rm BH}\sim10^{9}\,\rm M_{\odot}$ self-consistently in large-scale cosmological simulations require large simulated volumes, in line with new generation cosmological simulations \citep{2017MNRAS.467.4243D,2018MNRAS.473.4077P,2018arXiv181205609N}. Unfortunately, we will still crudely lack statistics with such simulations. 
Zoom-in simulations are the best way of simulating the evolution of BHs as massive as the observed quasars at $z\sim6$. 
Using this technique, \citet{Costa:2013aia} found that the mass of the simulated quasars were not quite exactly as massive as the most massive quasars observed at $z\sim 6$. They found the matching the observed quasars would require to introduce stronger galactic winds from supernovae. The winds coming out of the surrounding galaxies would push the gas toward the quasar host, therefore increasing its gas supply and consequently its growth \citep{Costa:2013aia}. 

In this paper, we find that the AGN feedback in Horizon-AGN does not have a strong impact on the galaxy number counts, at least with our modeling of the feedback. Similar results have been discussed in \citet{Costa:2013aia}.
However, from our analytical estimate we find that the size of the HII region around massive BHs is very anisotropic and could extend on potentially large scales. This could explain the absence of galaxies in the surroundings of some quasars \citep{2010ApJ...721.1680U}. Simulations with radiative feedback are needed to further understand how AGN feedback impact the surroundings of quasars. Radiative pressure-driven winds could also remove gas from the very nearby galaxies. 

In this work, we have found that there is also a pretty large variance in the star formation rates of the galaxies in the field of view of massive black holes. At least a few environments have high SFR, which would be similar to what has been found in the observed fields of high-redshift quasars if we scale the mass of our simulated galaxies to the mass of the observed companions. More detailed analyses of cosmological simulations are needed to understand the star-forming properties of massive BH galaxy companions, as a function of the stellar masses of these galaxies, their morphologies, and separation distances from the massive BH host. It will be crucial to understand the recent observations showing a high incidence of starburst galaxies in the field of view of quasars at high redshift. 

The present-day properties and environments of the descendants of the high-z quasars are still not well understood.
In a forthcoming paper we study the ability of the high-z environments, i.e. over-dense vs under-dense regions, to foster the subsequent growth of central supermassive BHs down to $z=0$.

\section*{Acknowledgements}
We thank Tiago Costa, Joe Silk, Christopher Hayward, Eric Gawiser, Jeff Newman, Steve Finkelstein, Nicolas Martinet, Clotilde Laigle, Aaron Yung and Alba Vidal-Garcia for fruitful discussions. 
MV acknowledges funding from the European Research Council under the European Community's Seventh Framework Programme (FP7/2007-2013 Grant Agreement no.\ 614199, project ``BLACK'').
RSS is grateful for the generous support of the Downsbrough family.
This work is partially supported by the Spin(e) grant {ANR-13-BS05-0005} (http://cosmicorigin.org) of the French {\sc Agence Nationale de la Recherche}, and ERC grant 670193.
CP thanks SUPA and the ERC for funding.
This work has made use of the Rusty Cluster at the Center for Computational Astrophysics (CCA) of the Flatiron Institute. The Flatiron Institute is supported by the Simons Foundation. This work also made use of the Horizon Cluster hosted by Institut d'Astrophysique de Paris (IAP), we thank Stephane Rouberol for running smoothly this cluster.

\bibliographystyle{mn2e}
\bibliography{biblio_complete,biblio,bib_chapter_reionization}

\appendix

\section{Observational evidence for overdense environment around high-redshift quasars?}
\label{sec:observations}
\subsection{Observations}
Over the last decade, an increasing number of observational studies have investigated galaxy overdensities around high-redshift quasars.
Here we summarize some of their findings. Of course, all the studies presented here use different fields of view, different criteria to select galaxies, different methods to identify these galaxies, and focus on different redshift in the range $z=4-7$.  We do not aim to perform an exhaustive comparison from one study to another (and in fact, several studies sometimes have different conclusions on the same quasar environment), but rather draw a global picture of the galaxy environment around high-redshift quasars, that we will use in the following sections to compare to the environment of simulated massive BHs from the large-scale simulations. Results from the different studies discussed below are shown in Fig.~\ref{fig:obs_data}. The figure shows counts of galaxies in the field of view of high-redshift quasars as a function of radial projected distance from these quasars.
To compute the projected distance we use the coordinates (right ascension and declination) of the objects that are given in the literature (the reference of the papers are listed on the right side of Fig.~\ref{fig:obs_data}). We compute the separation between the quasars and the different objects (galaxies or other quasars), in cMpc, using the cosmology of the simulation Horizon-AGN, described in the next section. 
Note that we show the number counts of galaxies, and not the excess of the number counts of galaxies compared to a reference number counts. This would have required to compute the individual theoretical function of the galaxy number counts in the absence of clustering.  This {\it non-clustering} function depends on distance from the quasar, but also the observed volume, and the observational selection of galaxies, and would therefore be different for all the observations listed below.
In Fig.~\ref{fig:obs_data}, we also show some of the fields of view that are probed by the instruments used in these studies (MUSE, HST ACS, GMOS-North, VLT FORS2, shown as vertical solid black lines). We also provide information regarding the instruments, and galaxy selection, for the different quasar fields in Table 1.

\begin{table*}
\caption{Summary of the sample of high-redshift quasars described in Section 2. We provide here the name of quasars, their redshift, reference of the study presenting the number counts of objects in their field of view, instruments and galaxy selection criteria.} 
\begin{center}
\begin{tabular}{ccccc}
\hline
\hline
Quasars & Redshift & References & Instruments & Galaxy selections  \\
\hline
J1439+0342 & 4.15 & Schneider+00 & & quasar pair  \\
J0256+0019 & 4.79 & McGreer+14 & ACS HST & $i, r, z, i, J, H$\\
J2130+0026 & 4.95 & Husband+13 & FORS2 VLT & $R,I,z$,$(R-I)>1$,$(I-z)<0.5$\\
J0221-0345 & 5.02 & McGreer+16 & CFHT & quasar pair\\
J0338+0021 & 5.03 & Husband+13 & FORS2 VLT & $R,I,z$, $(R-I)>1$, $(I-z)<0.5$\\
J1204+0021 & 5.09 & Husband+13 & FORS2 VLT & $R,I,z$, $(R-I)>1$, $(I-z)<0.5$\\
J0203+0012 & 5.72 & Banados+13 & FORS2 VLT & $(Z-NB)>0.75$, $(R-Z)>1$, $|(Z-NB)|>2.5\sqrt{\sigma^{2}_{\rm z}+\sigma^{2}_{\rm NB}}$\\
J2151-1604 & 5.73 & Mazzucchelli+16 & FORS2 VLT & $(Z-NB)>0.75$, $(R-Z)>1$, $|(Z-NB)|>2.5\sqrt{\sigma^{2}_{\rm z}+\sigma^{2}_{\rm NB}}$\\
J0836+0054 & 5.82 & Zheng+06 & ACS-HST & $i_{775}$, $z_{850}$, $z_{850}<26.5$, $(i_{775}-z_{850})<2$\\
J1306+0356 & 5.99 & Kim+09 & ACS HST & $i_{775}$, $z_{850}$, $(i_{775}-z_{850})>1.5$\\
J1630+4012 & 6.05 & Kim+09 & ACS HST & $i_{775}$, $z_{850}$, $(i_{775}-z_{850})>1.5$\\
 & & Willott+05 & GMOS-North & $(i'-z')>1.5$\\
 &  & Morselli+14 & Large Binocular Camera & $r,i,z$, $z_{AB}<25$, $i-z>1.4$\\
J0842+1218 & 6.08 & Decarli+17 & ALMA & \\
J2100+1715 & 6.08 & Decarli+17 & ALMA & \\
J1048+4637 & 6.23 & Kim+09 & ACS HST & $i_{775}$, $z_{850}$, $(i_{775}-z_{850})>1.5$\\
J303-21 & 6.08 & Decarli+17 & ALMA & \\
J0050+3445 & 6.25 & McGreer+14 & ACS HST & $i_{775}$, $Y_{105}$, $(i_{775}-Y_{105})>1.8$, $Y_{105}<25$\\
J1030+0524 & 6.28 & Kim+09 & ACS HST & $i_{775}$, $z_{850}$, $(i_{775}-z_{850})>1.5$\\
 & & Willott+05 & GMOS-North & $i'-z'>1.5$\\
 & & Stiavelli+05 & ACS HST & $i_{775}$, $z_{850}$, $(i_{775}-z_{850})\geqslant 1.5$, $z_{850}\leqslant 26.5$\\
 &  & Morselli+14 & Large Binocular Camera & $r,i,z$, $z_{AB}<25$, $i-z>1.4$\\
J1148+5241 & 6.4 & Kim+09 & ACS HST & $i_{775}$, $z_{850}$, $(i_{775}-z_{850})>1.5$\\
 & & Willott+05 & GMOS-North & $i'-z'>1.5$\\
 &  & Morselli+14 & Large Binocular Camera & $r,i,z$, $z_{AB}<25$, $i-z>1.4$\\
 J2329-0301 & 6.4 & Utsumi+10 & Subaru Suprime-Cam & $z_{R}, i', z'$, $(i'-z')>1.9$, $(z'-z_{R})>0.3$\\
 J2131-20 & 6.59 & Decarli+17 & ALMA & \\
 J0305-3150 & 6.61 & Farina+16 & MUSE VLT & \\
 J1120+0641 & 7.08 & Simpson+14 & ACS HST & $(i_{815}-Y_{105})>2.6$, $(Y_{105}-J_{125})<1$, $(i_{815}-Y_{105})>(Y_{105}-J_{125})$\\  
\hline
\end {tabular}
\end{center}
\label{obs_methods}
\end{table*}

First of all, \citet{2000AJ....120.2183S} report the discovering of a pair of quasars at $z=4.25$, including the quasar J143951+003429. 
The second quasar is identified at a separation of 33'', which correspond to $\sim 1.2 \, \rm cMpc$.
Another pair of quasars, including quasar J0221-0342, found at $z=5.02$ in the CFHT Legacy Survey, is described in \citet{2016AJ....151...61M}. They are separated by an angle of 21'', which correspond to a transverse separation of $\sim 0.81\, \rm cMpc$ at that redshift.
The discovery of these quasars pairs favors the idea of quasars being in overdense regions at high redshift and scale of $\sim \rm cMpc$.
Except these two special cases relating the discovery of quasar pairs, the other studies focus on the identification of galaxies around quasars. We report here several studies by chronological order of publication. 

\citet{2006ApJ...640..574Z} report the observation of 9 candidate galaxies around quasar SDSS J0836+0054, at $z=5.82$, with a minimum separation of $0.8\, \rm{cMpc}$.
This study of the overdense region uses the Advanced Camera for Surveys ACS at the Hubble Space Telescope (ACS HST).

The galaxy density around 5 high-redshift quasars is studied in \citet{2009ApJ...695..809K}, where half of the environments are underdense or similar to blank field, half of them are overdense. Quasar J1030+0524, at $z=6.28$, is close to 14 candidate galaxies. \citet{2005ApJ...622L...1S} also find evidence for an overdensity of LBGs in this quasar field using the ACS HST with a FoV of $11$ arcmin.
This quasar has been studied by \citet{Willotetal2005} with the GMOS-North imaging spectrograph with a FoV of $5'\times5'$, and has been studied recently by \citet{2014A&A...568A...1M} as well, with the large binocular Telescope, with a FoV of $8\times 8 \rm pMpc$.
Quasar J1630+4012 at $z=6.05$ \citep{2009ApJ...695..809K} is surrounded by 11 candidate galaxies, quasar J1048+4637 at $z=6.23$ by 7 candidate galaxies \citep[also studied by][]{Willotetal2005,2014A&A...568A...1M}, quasar J1148+5251 at $z=6.40$ by 3 candidate galaxies \citep[also studied by][]{Willotetal2005,2014A&A...568A...1M}, and finally one candidate galaxy is observed close to the last quasar J1306+0356 presented in \citet{2009ApJ...695..809K}. These numbers of candidate galaxies are upper limits.

\citet{2010ApJ...721.1680U} study the vicinity of the quasar CFHQS J2329-0301, at $z=6.4$, and find an overdensity of 7 LBG in the quasar field of view. These galaxies are concentrated in a ring-like shape, of radius $\sim 3\, \rm pMpc$, centered on the quasar. No overdensity is, however, found in the very close vicinity of the quasar. The strong UV radiation of the quasar could suppress galaxy formation in the close vicinity of the quasar, or the UV selection could fail to identify some galaxy with the UV spectrum being obscured by star formation. This study was the first to use a wide FoV of $34'\times27'$, with the Suprime-Cam at the Subaru Telescope.

The environment of ULAS J0203+0012 is studied in \citet{2013ApJ...773..178B}, this is a quasar at $z=5.72$, 2 candidate galaxies are found, but this is consistent with a blank field and therefore there is no evidence for an overdensity of galaxies around this quasar. For the first time, they search for LAE galaxies, for which a better estimate of the redshift can be achieved, with $\Delta z \sim 0.1$ (instead of $\Delta z \sim 1$ for LBGs). The observations have been conducted with the spectrograph FORS2 with a FoV of $6.8'\times6.8'$.

\citet{2013MNRAS.432.2869H} search for LBGs in 3 quasar fields, with FORS2 at the VLT. 7 LBGs and another quasar RD567 are identified around quasar J0338+0021 at $z=5.03$, 4 around quasar J1204-0021 at $z=5.09$, and 2 around quasar J2130+0026 at $z=4.95$. These numbers include only spectroscopically confirmed LBGs. Two of these 3 quasars show evidence for an enhancement in the counts. This work also argues that the studies based on imaging only may miss some of the objects on the high-redshift quasar fields, and that a control sample is necessary to evaluate the strength of the overdensities.

\citet{2014AJ....148...73M} study the environment of quasar CFHQS J0050+3445, observed at $z=6.25$ with the ACS HST and WFC3, one galaxy is observed, and could be at a separation of less than $\sim 5\, \rm{pkpc}$ from the quasar. The same work also relates the discovery of another galaxy, with a separation of less than $\sim 12 \, \rm{pkpc}$, in the field of quasar SDSS J025617.7+001904 at $z=4.79$. The close separation of these objects, could illustrate the mergers of two galaxies.

Using the LBG search technique and the ACS HST instrument, \citet{2014MNRAS.442.3454S} do not find any excess in the field of view of the most distant quasar ULAS J112001.48+064124.3 at $z=7.08$. 

In a recent paper, \citet{2017ApJ...834...83M} study the vicinities of the quasar PSO J215.1512-16.0417, searching for LAEs with FORS2 at the VLT, but also do not find any excess of galaxies.

However, \citet{2017ApJ...848...78F} recently related the discovery of a close LAE companion (12.5 pkpc separation) of the quasar J0305-3150, observed at $z=6.61$. This study has been carried with the Multi Unit Spectroscopic Explorer MUSE at the VLT.

\citet{2017Natur.545..457D} describe the recent discovery of a companion star-forming galaxy in the field of view of four quasars with the Atacama Large Millimeter Array (ALMA): quasar SDSS J0842+1218 ($z=6.08$, projected separation of 47.7 pkpc), quasar CFHQ J2100-1715 ($z=6.08$, projected separation of 60.7 pkpc), quasar PSO J231-20 ($z=6.59$, projected separation of 8.4 pkpc), and quasar PSO J303-21 (at $z=6.23$, projected separation of 13.8 pkpc). For the first time, this identifies galaxies that are forming stars very efficiently at high redshift, more than $100\, \rm M_{\odot}/yr$, while so far only quasar host galaxies were known to be able to do so.

\subsection{Discussion on the uncertainties}
As the enhancement of galaxy counts around high-redshift quasars is not clear, in the following we summarize the main reasons that could lead to such variability/discrepancy in the findings of all these studies. First of all, searching for LBG galaxies at high redshift of $z=5-6$ is very challenging, because of both the uncertainties in the number counts of dropout for LBG galaxies, and the large uncertainties regarding the redshift of those galaxies in the quasar fields.
Recently, the search for LAE galaxies have been used as another method to study high-redshift quasar environment, this is an important improvement, because reducing the redshift uncertainty of the galaxies from $\Delta z\sim 1$ for LBGs to $\Delta z\sim 0.1$ for LAEs. 
Some of the studies presented below could suffer from probing only relatively small scale environment around the quasars. In fact, some quasars have been observed several times, with different instruments, selection criteria, and FoV. This is the case for the quasars J1148+5251 \citep{Willotetal2005, 2009ApJ...695..809K,2014A&A...568A...1M}, J1048+4637 \citep{Willotetal2005,2009ApJ...695..809K,2014A&A...568A...1M}, and J1030+0524 \citep{2005ApJ...622L...1S,Willotetal2005,2009ApJ...695..809K,2014A&A...568A...1M}, as mentioned in the description above. These different studies yield different findings, and particularly larger FoVs seem to show enhancements of galaxies, that were not found with previous smaller FoV studies. The idea of observing the enhancement of galaxies on larger scales was triggered by \citet{2009MNRAS.394..577O}, which show that the enhancement of galaxies can be on scales as large as $70 \, \rm cMpc$ at $z=6$, that are too large to be probed by HST ACS, or GMOS-North, for example. Also, the enhancement at smaller scales, could be attenuated by AGN feedback in the HII region (sphere of ionized gas) around massive BHs, which may be able to cease star formation in the surrounding galaxies \citep{2007ApJ...663..765K,2010ApJ...721.1680U}. We address this point in more detail in Section 5.2.

The counts of LAEs could also suffer from the complicated radiative transfer of the resonant Lyman $\alpha$ emission line and a large scattering cross-section, which could generate anisotropic selection biais \citep{2011ApJ...726...38Z}. The consequent observed Ly$\alpha$ emission is very sensitive to the local environment properties, such as matter density, line-of-sight velocity, density gradient on the line-of-sight, and the velocity gradient on the line-of-sight and perpendicular to it. 
For example, the probability of observing LAEs is higher in under dense regions than in overdense regions if density fluctuations are along the line-of-sight, but the probability becomes higher in over dense regions if the fluctuations are transverse to the line-of-sight. This results in an attenuation of the line-of-sight density fluctuations and an enhancement of transverse density fluctuations.
We also note that these effects have been recently investigated with the Illustris simulations, and are found to be smaller \citep{2017arXiv171006171B}. 

There are many uncertainties in the measurement of the galaxy overdensity traced by number counts around high-redshift quasars. Using deeper, and multi-wavelength imaging, as well as spectroscopic confirmation could help to detect galaxies that could have been missed by the current galaxy selections, such as the very faint galaxies, the star-forming galaxies that are strongly obscured, and the galaxies without star forming activity. The upcoming James Webb Space Telescope (JWST) could be able to further investigate the presence of these galaxies in quasar fields of view in the near future. Also, using large fields of view \citep{2009MNRAS.394..577O,2010ApJ...721.1680U,2014A&A...568A...1M} could allow us to detect larger scale overdensity, that could be missed by other studies using smaller FoV. Using additional control fields would increase our understanding of the definition of overdensity.
In the following we use two state-of-the-art simulations Horizon-AGN and Horizon-noAGN to investigate the number counts of galaxies around massive BHs.

\section{Reference galaxy number counts for randomly distributed galaxies}
\label{sec:number_count}
\begin{figure*}
\includegraphics[width=\textwidth]{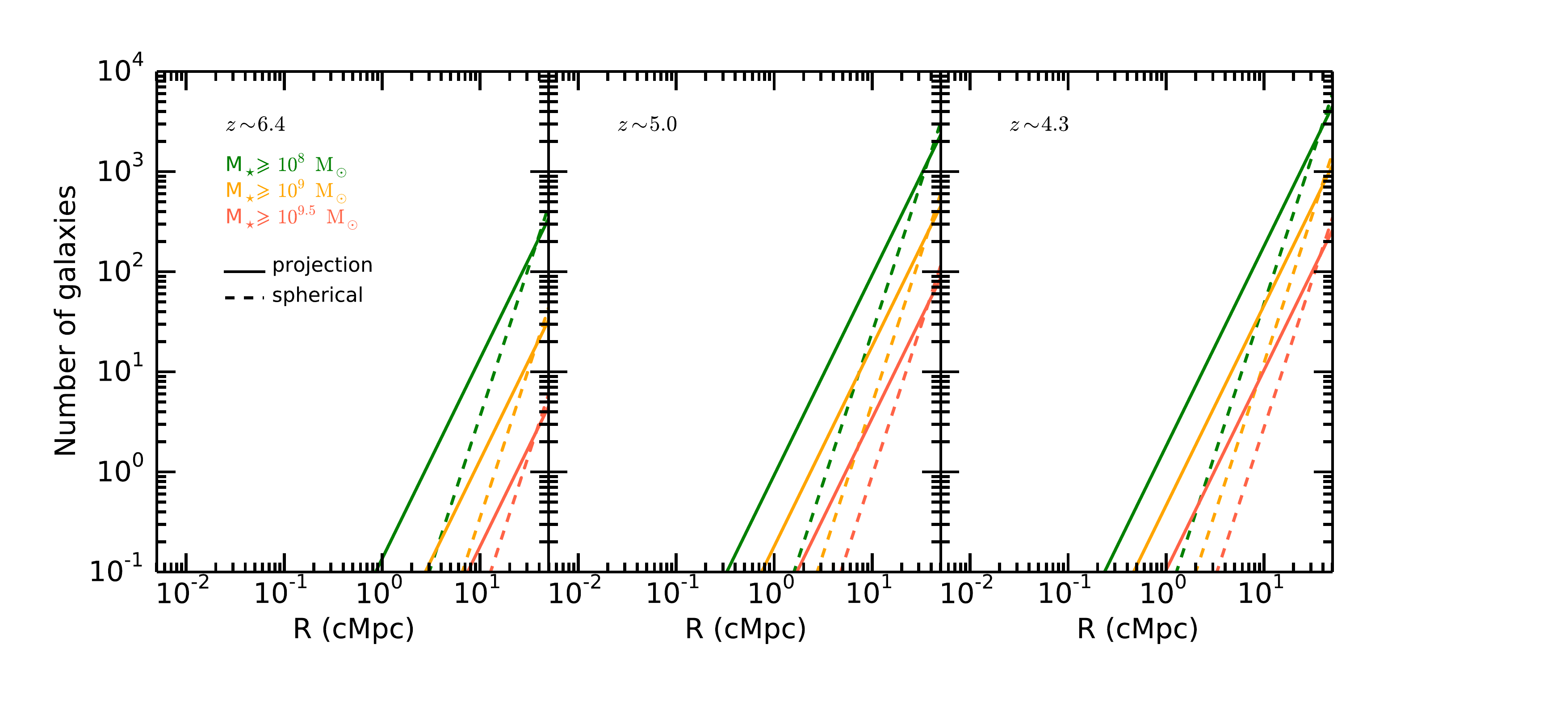}
\caption{Number of galaxy counts in the absence of clustering, for different cuts in galaxy stellar mass (green lines: $\rm M_{\star, thres}=10^{8}\, M_{\odot}$; orange lines: $\rm M_{\star, thres}=10^{9}\, M_{\odot}$; red lines: $\rm M_{\star, thres}=10^{9.5}\, L_{\odot}$), and for spherical radii (dashed lines) or projected separation distance between galaxies (solid lines).}
\label{fig:reference_noclus}
\end{figure*}

We show the number counts of galaxies that we expect in the simulation for a non-clustered region in Fig.~\ref{fig:reference_noclus}. We show these functions for different cuts in luminosity (green for $\rm M_{\star, thres}=10^{8}\, M_{\odot}$; orange for $\rm M_{\star, thres}=10^{9}\, M_{\odot}$; red for $\rm M_{\star, thres}=10^{9.5}\, M_{\odot}$), and for spherical radii (dashed lines) or projected separation distance between galaxies (solid lines).

\section{Galaxy number counts and dark matter halo mass}
\label{sec:haloes}
\begin{figure}
\includegraphics[scale=0.5]{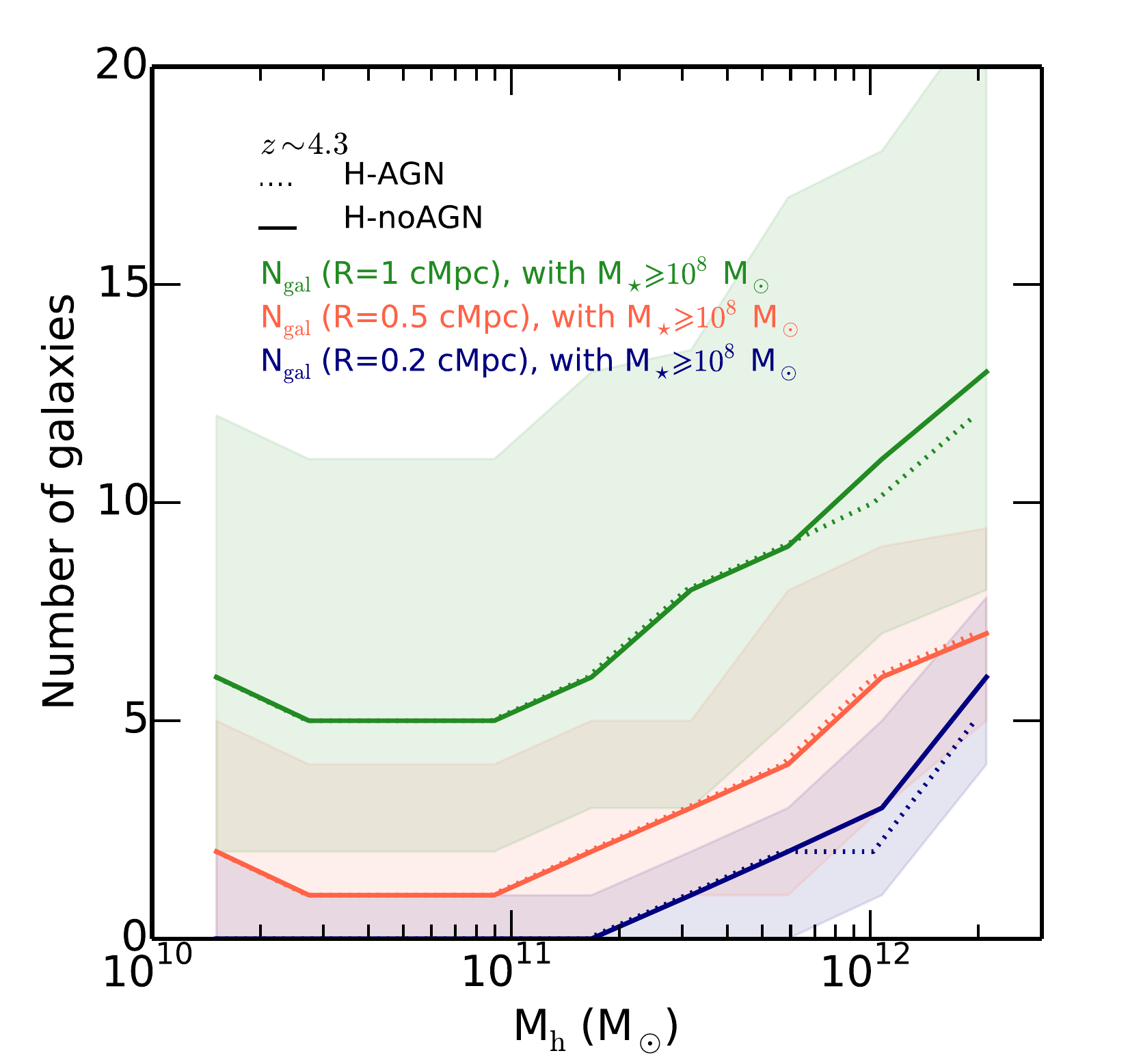}
\caption{Median of the galaxy number counts at a given projected distance $R$ in the field of view of galaxies as a function of their dark matter halo mass, for the simulation Horizon-noAGN (solid lines), and Horizon-AGN (dotted lines). Here we consider a stellar mass cut of $M_{\rm \star, thres}\geqslant 10^{8}\, \rm M_{\odot}$, and a projected distance of $50\, \rm cMpc$, but similar results are found for different stellar mass cuts, and projected distance of 10 cMpc.}
\label{fig:haloes_HnoAGN}
\end{figure}

In Fig.~\ref{fig:haloes_HnoAGN}, we show the median of the galaxy number counts at a given projected distance $R$ in the field of view of galaxies as a function of their dark matter halo mass, for the simulation Horizon-noAGN (solid lines), and Horizon-AGN (dotted lines). We find very similar correlations, with a slightly higher number counts of galaxies for the simulation with no AGN feedback for the most massive haloes.

\begin{figure*}
\centering
\includegraphics[scale=0.53]{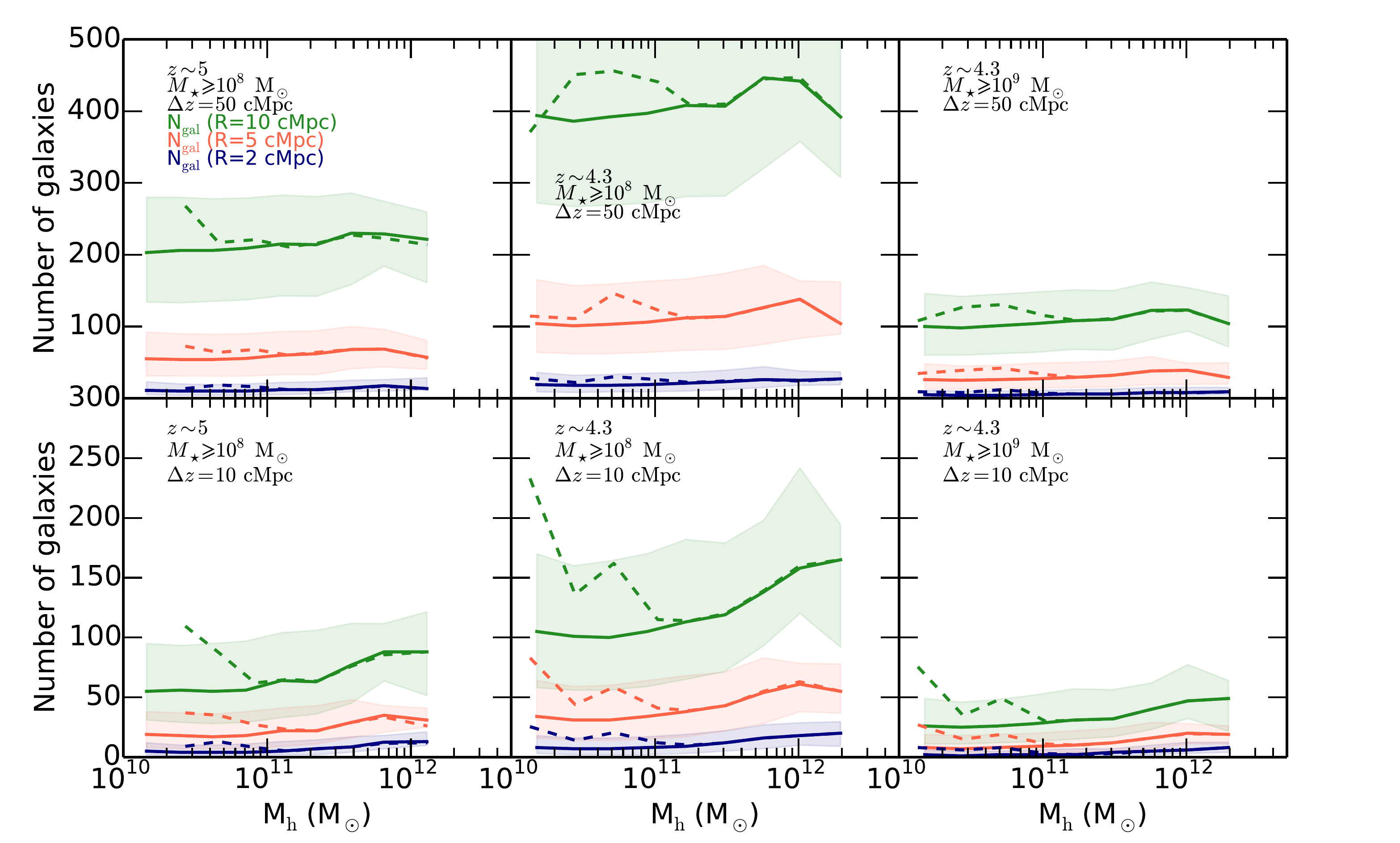}
\caption{Median of the galaxy number counts at a given projected distance $R$ in the field of view of galaxies as a function of their dark matter halo mass. Same as Fig.~\ref{fig:halo} for higher apertures of $R=2, 5, 10 \, \rm cMpc$ in blue, red and green, respectively.}
\label{fig:halo_higher_apertures}
\end{figure*}

\section{Profiles along lines-of-sight}
\label{sec:profiles}
\begin{figure}
\includegraphics[scale=0.58]{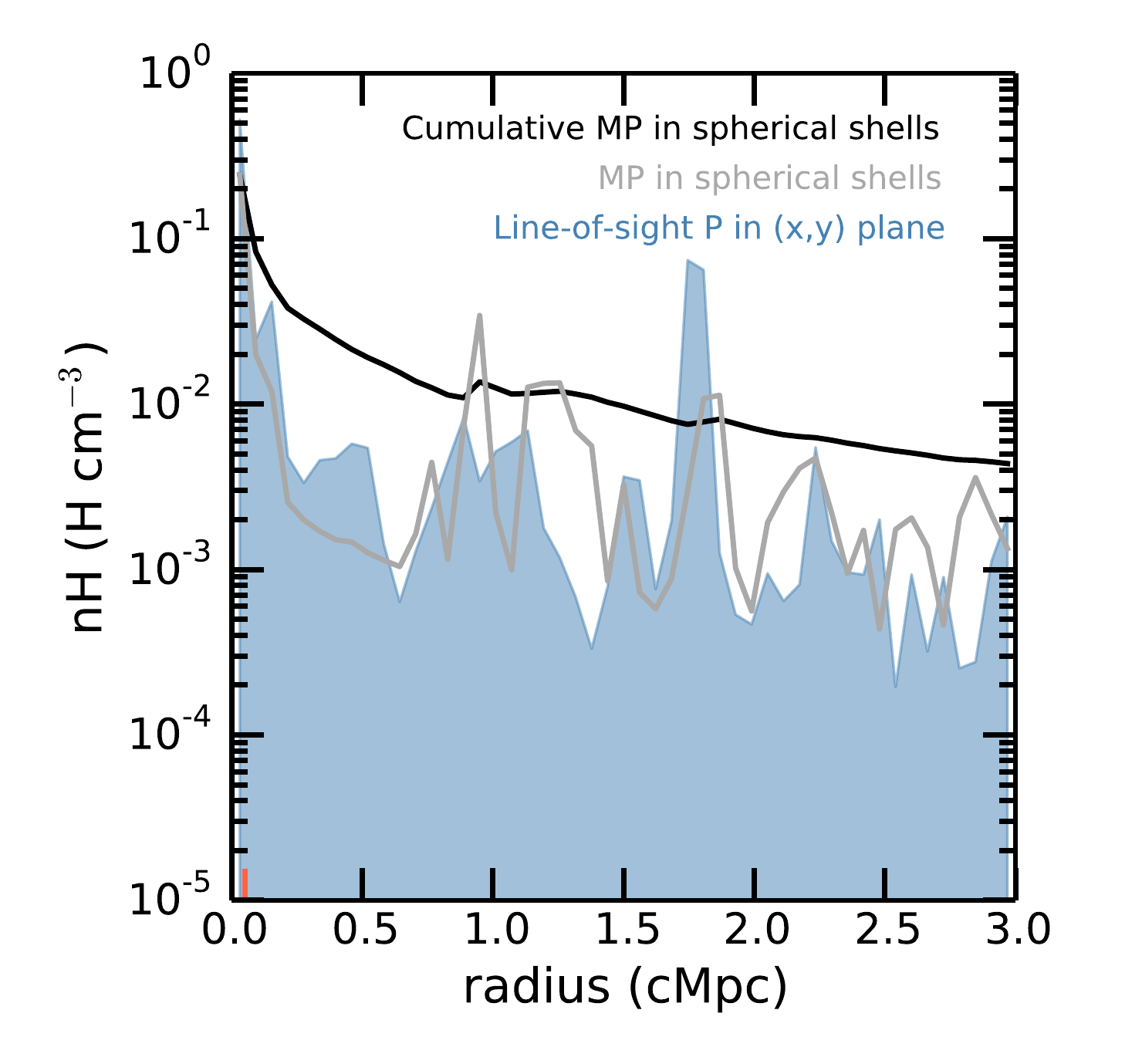}
\caption{Minimum and maximum values of $n_{\rm H}$ profiles (blue shaded area) along 40 lines-of-sight from an individual BH in the plane (x,y). For comparison we show with the grey line the 3D profile, computed in spherical shells around the BH. We also show the cumulative profile with the black line (where for each radius the $n_{\rm H}$ value encloses all the $n_{\rm H}$ content from the galaxy to the given radius).
Finally, the red symbol on the bottom left of the figure indicates the virial radius of the BH host galaxy. }
\label{fig:profile}
\end{figure}

In Fig.\ref{fig:profile} we show in blue shaded area the minimum and maximum of the $n_{\rm H}$ profiles along the 40 lines-of-sight from an individual BH. These lines-of-sight are all in the plane (x,y). For comparison we also show in grey line the mean $n_{\rm H}$ profile computed in spherical shells around the BH. The black line indicates the cumulative $n_{\rm H}$ profile. The little red symbol on the bottom left of the figure indicates the virial radius of the BH host galaxy. We use the $n_{\rm H}$ profiles along the lines-of-sight to compute the shape and size of the HII region in the plane (x,y) around this BH.

\end{document}